\documentclass[12pt]{article}
\usepackage{epsfig}
\usepackage{wrapfig}
\usepackage{subfigure}
\usepackage{amsmath}
\usepackage{hhline}
\usepackage{amssymb}
\usepackage{times}
\usepackage{cite}

\newlength{\dinwidth}
\newlength{\dinmargin}
\begin{document}  
% The rest
%%%%%%%%%%%%%%%%%%%%%%%%%%%%%%%%%%%%%%%%%%%%%%%%%%%%%%%%%%%%%%%%%%
%%%%%%%%%%%%%%%%%%     command abbreviations    %%%%%%%%%%%%%%%%%%
%%%%%%%%%%%%%%%%%%%%%%%%%%%%%%%%%%%%%%%%%%%%%%%%%%%%%%%%%%%%%%%%%%
\newcommand{\be}{\begin{equation}}
\newcommand{\ee}{\end{equation}}
\newcommand{\bea}{\begin{eqnarray}}
\newcommand{\eea}{\end{eqnarray}}
\newcommand{\nn}{\nonumber}
\newcommand{\sr}{\stackrel}
\newcommand{\D}{\displaystyle}
\newcommand{\half}{\textstyle{\frac{1}{2}}}
\newcommand{\bs}{\boldsymbol}

\newcommand{\pom}{{I\!\!P}}
\newcommand{\reg}{{I\!\!R}}
\newcommand{\gap}{\stackrel{>}{\sim}}
\newcommand{\lap}{\stackrel{<}{\sim}}

\newcommand{\alps}{\alpha_s}
\newcommand{\sqrts}{$\sqrt{s}$}
\newcommand{\LO}{$O(\alpha_s^0)$}
\newcommand{\Oa}{$O(\alpha_s)$}
\newcommand{\Oaa}{$O(\alpha_s^2)$}
\newcommand{\PT}{p_{\perp}}
\newcommand{\PO}{I\!\!P}
\newcommand{\xpomlo}{3\times10^{-4}}  
\newcommand{\dgr}{^\circ}
\newcommand{\pbarnt}{\,\mbox{{\rm pb$^{-1}$}}}

% 
% DVCS KIN
%
\newcommand{\tg}{\theta_{\gamma}}
\newcommand{\te}{\theta_e}
%
% Some useful tex commands
%
% kinematic variable
\newcommand{\qsq}{\mbox{$Q^2$}}
\newcommand{\Qsq}{\mbox{$Q^2$}}
\newcommand{\s}{\mbox{$s$}}
\newcommand{\ttra}{\mbox{$t$}}
\newcommand{\modt}{\mbox{$|t|$}}
\newcommand{\eminpz}{\mbox{$E-p_z$}}
\newcommand{\eminpzs}{\mbox{$\Sigma(E-p_z)$}}
\newcommand{\rap}{\ensuremath{\eta^*} }
\newcommand{\W}{\mbox{$W$}}
\newcommand{\w}{\mbox{$W$}}
\newcommand{\Q}{\mbox{$Q$}}
\newcommand{\q}{\mbox{$Q$}}
\newcommand{\xB}{\mbox{$x$}}  % Bjorken x
\newcommand{\xF}{\mbox{$x_F$}}  % Feynman x
\newcommand{\xg}{\mbox{$x_g$}}  % x_g
\newcommand{\xbj}{x}
\newcommand{\xpom}{x_{\PO}}
\newcommand{\y}{\mbox{$y~$}}

\newcommand{\gp}{\ensuremath{\gamma^*}p }
\newcommand{\gammasp}{\ensuremath{\gamma}*p }
\newcommand{\gammap}{\ensuremath{\gamma}p }
\newcommand{\gsp}{\ensuremath{\gamma^*}p }
% ep
\newcommand{\epem}{\mbox{$e^+e^-$}}
\newcommand{\ep}{\mbox{$ep~$}}
\newcommand{\epl}{\mbox{$e^{+}$}}
\newcommand{\emi}{\mbox{$e^{-}$}}
\newcommand{\epm}{\mbox{$e^{\pm}$}}
%=====================================================================
% elastic VM abbreviations
%
% VM
\newcommand{\photon}{\mbox{$\gamma$}}
\newcommand{\phib}{\mbox{$\varphi$}}
\newcommand{\rh}{\mbox{$\rho$}}
\newcommand{\rhz}{\mbox{$\rh^0$}}
\newcommand{\ph}{\mbox{$\phi$}}
\newcommand{\om}{\mbox{$\omega$}}
\newcommand{\ome}{\mbox{$\omega$}}
\newcommand{\jpsi}{\mbox{$J/\psi$}}
\newcommand{\JPSI}{J/\psi}
\newcommand{\ups}{\mbox{$\Upsilon$}}
% b parameter
\newcommand{\bsl}{\mbox{$b$}}
%
%=====================================================================
% units
%
\newcommand{\cm}{\mbox{\rm cm}}
\newcommand{\GeV}{\mbox{\rm GeV}}
\newcommand{\gev}{\mbox{\rm GeV}}
\newcommand{\GeVx}{\rm GeV}
\newcommand{\gevx}{\rm GeV}
\newcommand{\MeV}{\mbox{\rm MeV}}
\newcommand{\mev}{\mbox{\rm MeV}}
\newcommand{\MeVx}{\mbox{\rm MeV}}
\newcommand{\mevx}{\mbox{\rm MeV}}
\newcommand{\GeVsq}{\mbox{${\rm GeV}^2$}}
\newcommand{\gevsq}{\mbox{${\rm GeV}^2$}}
\newcommand{\gevsqc}{\mbox{${\rm GeV^2/c^4}$}}
\newcommand{\gevcsq}{\mbox{${\rm GeV/c^2}$}}
\newcommand{\mevcsq}{\mbox{${\rm MeV/c^2}$}}
\newcommand{\GeVsqm}{\mbox{${\rm GeV}^{-2}$}}
\newcommand{\gevsqm}{\mbox{${\rm GeV}^{-2}$}}
\newcommand{\nb}{\mbox{${\rm nb}$}}
\newcommand{\nbinv}{\mbox{${\rm nb^{-1}}$}}
\newcommand{\pbinv}{\mbox{${\rm pb^{-1}}$}}
\newcommand{\mm}{\mbox{$\cdot 10^{-2}$}}
\newcommand{\mmm}{\mbox{$\cdot 10^{-3}$}}
\newcommand{\mmmm}{\mbox{$\cdot 10^{-4}$}}
\newcommand{\degr}{\mbox{$^{\circ}$}}
%
%=====================================================================
% inequation symbols
%
%greater than or approx. symbol
\def\gsim{\,\lower.25ex\hbox{$\scriptstyle\sim$}\kern-1.30ex%
  \raise 0.55ex\hbox{$\scriptstyle >$}\,}
%less than or approx. symbol
\def\lsim{\,\lower.25ex\hbox{$\scriptstyle\sim$}\kern-1.30ex%
  \raise 0.55ex\hbox{$\scriptstyle <$}\,}
%
%
%=====================================================================

%\linenumbers

\begin{titlepage}

\begin{flushleft}
DESY 07-142 \hfill ISSN 0418-9833 \\
September 2007
\end{flushleft}

\vspace{2cm}

\begin{center}
\begin{Large}

{\bf Measurement of Deeply Virtual Compton Scattering \\
and its $t$-dependence at HERA }

\vspace{2cm}

H1 Collaboration

\end{Large}
\end{center}

\vspace{2cm}

%=========================================================================
\begin{abstract}

A measurement of elastic deeply virtual Compton 
scattering $\gamma^* p \rightarrow \photon p$ using  $e^- p$ collision data
recorded with the H1 detector at HERA is presented. 
The analysed data sample corresponds to 
an integrated luminosity of $145$ pb$^{-1}$.
The cross section is measured as a function of the virtuality
$Q^2$ of the exchanged photon
and the centre-of-mass energy $W$ of the $\gamma^*p$ system 
in the kinematic domain 
$6.5 < Q^2 < 80$~GeV$^2$, $30 < W < 140$~GeV and $|t| < 1$ GeV$^2$,
where  $t$ denotes the  squared momentum transfer at the proton vertex.
The cross section is determined differentially in $t$ 
for different $Q^2$ and $W$ values and
exponential $t$-slope parameters are derived. 
The measurements are compared to a NLO QCD calculation based on generalised 
 parton distributions.
In the context of the dipole approach, the geometric scaling property of the DVCS cross section 
\mbox{is studied for different values of $t$.}

\end{abstract}
%=========================================================================

\vspace{1.5cm}

\begin{center}
Submitted to Phys. Lett. {\bf B}
\end{center}

\end{titlepage}

\begin{flushleft}
%-- H1AUTS Author list by names 
%-- Status: Tue Jul 10 09:40:42 CEST 2007  Number of authors = 288 

F.D.~Aaron$^{5,49}$,           %BUCH-PD        11/06           Aaron               
A.~Aktas$^{11}$,               %DESY-LEFT      09/06           Aktas               
C.~Alexa$^{5}$,                %BUCH-PD        06/06           Alexa               
V.~Andreev$^{25}$,             %LPI -PD        8/88            Andreev             
B.~Antunovic$^{11}$,           %DESY-PD        05/07           Antunovic           
S.~Aplin$^{11}$,               %DESY-PD        01/04           Aplin               
A.~Asmone$^{33}$,              %ROME-ST        07/2            Asmone              
A.~Astvatsatourov$^{4}$,       %BRUX-PD        07/04           Astvatsatourov      
S.~Backovic$^{30}$,            %PODG-PD        03/2            Backovic            
A.~Baghdasaryan$^{38}$
% ...

%-- H1 Institutes 
\bigskip{\it
 $ ^{1}$ I. Physikalisches Institut der RWTH, Aachen, Germany$^{ a}$ \\
 $ ^{2}$ Vinca  Institute of Nuclear Sciences, Belgrade, Serbia \\
% ...
$ ^{4}$ Inter-University Institute for High Energies ULB-VUB, Brussels;
          Universiteit Antwerpen, Antwerpen; Belgium$^{ c}$ \\
 $ ^{5}$ National Institute for Physics and Nuclear Engineering (NIPNE) ,
          Bucharest, Romania \\
% ...
$ ^{11}$ DESY, Hamburg, Germany \\
% ...
$ ^{25}$ Lebedev Physical Institute, Moscow, Russia$^{ e}$ \\
% ...
$ ^{30}$ Faculty of Science, University of Montenegro,
          Podgorica, Montenegro$^{ e}$ \\
% ...
$ ^{33}$ Dipartimento di Fisica Universit\`a di Roma Tre
          and INFN Roma~3, Roma, Italy \\
% ...
$ ^{38}$ Yerevan Physics Institute, Yerevan, Armenia \\
% ...
$ ^{49}$ Also at Faculty of Physics, University of Bucharest,
          Bucharest, Romania \\
% ...
$ ^a$ Supported by the Bundesministerium f\"ur Bildung und Forschung, FRG,
      under contract numbers 05 H1 1GUA /1, 05 H1 1PAA /1, 05 H1 1PAB /9,
      05 H1 1PEA /6, 05 H1 1VHA /7 and 05 H1 1VHB /5 \\
% ...
$ ^c$ Supported by FNRS-FWO-Vlaanderen, IISN-IIKW and IWT
      and  by Interuniversity
Attraction Poles Programme,
      Belgian Science Policy \\
% ...
$ ^e$ Supported by the Deutsche Forschungsgemeinschaft \\
% ...
}
\end{flushleft}

\newpage

%=========================================================================
\section{Introduction}
%=========================================================================
\label{sec:intro}

Measurements of inclusive deep-inelastic scattering (DIS) of leptons and nucleons
allow the extraction of Parton Distribution Functions (PDFs) which describe
the fraction of the longitudinal momentum of the nucleon carried by the quarks, anti-quarks and gluons.
A shortfall of this approach is that the PDFs
contain information neither on the 
correlations between partons nor on their transverse distributions. 
This missing
information can be provided by measurements of processes 
in which the
nucleon remains intact and the 
four momentum transfer squared at the nucleon vertex, $t$, is non-zero~\cite{bernard,bukpaper, diehlpaper,mueller,guzey,strikpaper}.
The simplest such reaction is deeply virtual
Compton scattering (DVCS), the diffractive
scattering of a virtual photon off a proton
$\gamma^* p \rightarrow \gamma p$. 
In high energy electron-proton collisions at HERA,
DVCS is accessed through the reaction $e  p \rightarrow e  \photon  p$~\cite{Adloff:2001cn,dvcsh1,dvcszeus}.
This reaction also receives a contribution from the purely 
electromagnetic Bethe-Heitler (BH) process, where the photon is emitted from the electron. 
The BH cross section is precisely calculable in QED
and can be subtracted from the total process rate to extract
the DVCS cross section.

Perturbative QCD calculations assume that the DVCS reaction involves two 
partons in the proton which
carry different longitudinal and transverse momenta.
The difference in longitudinal momentum of the two involved partons, also called skewing, is a consequence of the mass
difference between the incoming virtual photon and the outgoing real
photon.
The skewing can be described by introducing
generalised parton distributions (GPDs) \cite{bernard,bukpaper,diehlpaper,mueller,guzey}, which are functions
of the two unequal momenta and thus encode information on the 
longitudinal momentum correlations of partons.
Information on the transverse momentum of partons is incorporated in the $t$-dependence of GPDs~\cite{bukpaper,diehlpaper,mueller,guzey}.
The $t$-dependent functions follow particular equations for their evolution
as a function of the four momentum transfer squared $Q^2$ of the 
exchanged virtual photon~\cite{diehlpaper,mueller,guzey}. These evolution equations still need to be tested.

The DVCS cross section can also be interpreted within the dipole model~\cite{dd,favart,lolo}.
In this picture the virtual photon fluctuates into a colour singlet 
$q\bar q$ pair (or dipole) of a transverse size $r\!\sim\!1/Q$,
which subsequently undergoes hard scattering with the gluons in the
proton~\cite{dipole}.
At very small values of the Bjorken scaling variable $x$
the saturation regime of QCD can be reached.
In this domain, the gluon density in the proton is so large that 
non-linear effects like gluon recombination tame its growth.
In the dipole model approach, the transition to the saturation regime is characterised by the so-called 
saturation scale parametrised here as $Q_s(x)=Q_0 ({{x_0}/{x}})^{-\lambda/2}$, where $Q_0$, $x_0$ and $\lambda$ are parameters~\cite{iim2}. 
The transition to saturation occurs when $Q$ becomes comparable to 
$Q_s(x)$.
An important feature of dipole models that incorporate saturation is that the total cross section can be expressed as a function of the single variable $\tau$:

\begin{equation}
\sigma_{tot}^{\gamma^\ast p}(x,Q^2)  = \sigma_{tot}^{\gamma^\ast p}( \tau ) , \;\;  \mbox{with} \; \ \ \ \tau=\frac{Q^2}{Q_s^2(x)}.
\label{eq:tau}
\end{equation}
This property, called geometric scaling, has already been observed  to hold
for the total $ep$ DIS cross section~\cite{golec,Stasto:2000er} as well as in DIS on nuclear targets~\cite{Freund:2002ux} and in diffractive processes~\cite{lolo}. 
It has also recently been addressed in the context of
exclusive processes including DVCS~\cite{lolo} and extended to cases with non-zero momentum transfer to the proton~\cite{robilast}.

This paper presents a new measurement of single and double differential DVCS cross sections as a function of
$Q^2$ and the $\gamma^*p$ centre-of-mass energy $W$. The single differential cross section $d\sigma / dt$ is also extracted. 
The data were recorded in the years $2005$ and $2006$ with the H1 detector
when HERA collided protons of $920$~GeV energy with
$27.6$~GeV electrons.
The sample corresponds to an integrated luminosity of  $145$ pb$^{-1}$,
four times larger than the previous H1 measurement \cite{dvcsh1}
of DVCS in positron-proton collisions.
The measurement is carried out 
in the kinematic range
$6.5 < Q^2 < 80$~GeV$^2$, $30 < W < 140$~GeV 
and $|t| <$ 1 GeV$^2$.
The $t$-dependence of the
DVCS cross section, $d\sigma / dt$, is found to be well approximated by an exponential form $e^{-b|t|}$; this parametrisation is used throughout the paper.
The $Q^2$ and $W$ dependences of $b$ are studied.
A parametrisation of the observed $Q^2$ dependence of $b$
is used to constrain the normalisation of the
pQCD predictions based on GPDs. The validity of the skewed evolution equations is tested. 
The geometric scaling property of DVCS is also investigated and the cross section is compared
with dipole model predictions.
The scaling property is studied for the first time for different values of $t$.

%

%=========================================================================
\section{Experimental Conditions and Monte Carlo Simulation} \label{simul}
%=========================================================================

A detailed description of the H1 detector can be found in~\cite{h1dect}.
Here, only the detector components relevant for the present analysis are
described. 
H1 uses a right-handed coordinate system with the $z$ axis along
the beam direction, the $+z$ or ``forward'' direction being that of the outgoing proton beam.
The polar angle $\theta$ is defined with respect to the $z$ axis and the
pseudo-rapidity is given by $\eta=-\ln \tan \theta /2$. 
The SpaCal~\cite{Appuhn:1996na}, a lead scintillating fibre calorimeter, 
covers the backward 
region ($153 ^{\rm \circ} < \theta < 176 ^{\rm \circ}$).
Its energy resolution for electromagnetic showers is $\sigma(E)/E
\simeq 7.1\%/\sqrt{E/{\rm GeV}} \oplus 1\%$. 
The liquid argon (LAr) calorimeter ($4^{\rm \circ} \leq \theta \leq
154^{\rm \circ}$) is situated inside a solenoidal magnet. 
The energy resolution for electromagnetic showers is 
$\sigma(E)/E \simeq 11\%/\sqrt{E/{\rm GeV}}$ as obtained from test beam 
measurements~\cite{Andrieu:1994yn}.
The main component of the central tracking detector is the central jet
chamber CJC ($20^{\rm \circ} < \theta < 160^{\rm \circ}$) which consists of 
two coaxial cylindrical drift chambers
with wires parallel to the beam direction.
The measurement of charged particle transverse momenta is performed
in a magnetic field of $1.16$~T, which is uniform over the full tracker volume.
The innermost proportional chamber CIP ($9^\circ < \theta < 171^\circ$)
is used in this analysis to complement the CJC in the backward region for the reconstruction of the interaction
vertex.
The forward muon detector (FMD) consists of
a series of drift chambers covering the range $1.9<\eta<3.7$. 
Primary particles produced at larger $\eta$ can be detected indirectly in the FMD if they undergo a secondary scattering with the beam pipe or other adjacent material. 
Therefore, the FMD is used in this analysis to provide an additional veto against inelastic or proton dissociative events.
The luminosity is determined from the rate of Bethe-Heitler processes
measured using a calorimeter located
close to the beam pipe at $z=-103~{\rm m}$ in the backward direction.

A dedicated event trigger was set up for this analysis.
It is based on topological and neural network algorithms
and uses correlations between electromagnetic energy 
deposits of electrons or photons in both the LAr and the SpaCal~\cite{roland}.
The combined trigger efficiency is close to $100$\%.

Monte Carlo (MC) simulations are used to estimate the background contributions and the corrections that must be applied to the data to account for the finite acceptance and the resolution of the detectors.
Elastic DVCS events in $ep$ collisions are generated using the Monte Carlo generator
MILOU~\cite{milou}, based on the cross section calculation from~\cite{ffspaper} and  using a $t$-slope parameter $b=5.45$~GeV$^{-2}$, as determined in this analysis (see section~\ref{sec:results}).
Inelastic DVCS events in which the proton dissociates into a baryonic system $Y$ are
also simulated with MILOU setting the $t$-slope
$b_{pdiss}$ to $1.2$~GeV$^{-2}$, as determined in a dedicated study (see section~\ref{selection}).
The Monte Carlo program COMPTON~2.0~\cite{compton2} is used to
simulate elastic and inelastic BH events.
The background source of
diffractive  meson events is simulated using the
DIFFVM Monte Carlo~\cite{diffvm}.
All generated events are passed through a detailed simulation of the H1
detector and are subject to the same reconstruction and analysis chain
as the data.

%=========================================================================
\section{Event Selection}  \label{selection}
%=========================================================================
%

In elastic DVCS events, the scattered electron and the photon
are the only particles that should give signals in the detector~\cite{dvcsh1}.
The scattered proton escapes undetected through the beam pipe.
The selection of the analysis event sample requires the scattered electron to be detected in the SpaCal and the photon in the LAr.
The energy of the scattered electron candidate must be greater than $15$ GeV.
The photon is required to have
a transverse momentum $P_T$ above $2$ GeV 
and a polar angle between $25^{\circ}$ and $145^\circ$.
Events are selected if there are either no tracks at all or a single central track which is associated with the scattered electron.
In order to reject inelastic and proton dissociation events,
no further energy deposit in the LAr calorimeter larger than
$1$~GeV is allowed and no activity above
the noise level should be present in the FMD.
The influence of QED radiative corrections is reduced by the requirement that
the longitudinal momentum balance $E - P_z$ be greater than $45$~GeV.
Here, $E$ denotes the energy and $P_z$ the momentum along the beam axis of all measured final state particles.
To enhance the DVCS signal with respect to the BH
contribution and to ensure a large acceptance, the
kinematic domain is  restricted to
$6.5<Q^2<80 $ GeV$^2$ and
$30<W<140 $~GeV. 

The selected analysis sample contains $2538$ events.
It is dominated by elastic DVCS events, 
but also contains contributions from the elastic
BH process and from the BH and DVCS processes with proton
dissociation, $ e^-  p \rightarrow e^-  \photon  Y$,
where the baryonic system $Y$ of mass $M_Y$ is undetected.
These background contributions are studied in further detail.
A control sample of BH events is selected. 
For this sample, it is required that the
electron be detected in the LAr and the photon in the SpaCal (see~\cite{dvcsh1} for more details).
The COMPTON MC describes accurately the
normalisation and the shapes of the distributions of the kinematic
variables for these events. 
The deviations are within $3$\%, and this value is used subsequently
as an estimate for the systematic uncertainty 
on this contribution.
A second control sample dominated by
inelastic BH and DVCS processes is obtained
by selecting events with a signal in the FMD. 
After subtracting the inelastic BH contribution, 
as estimated from the COMPTON MC,
this sample allows the normalisation of the inelastic DVCS process to be determined.
Within the model used in MILOU~\cite{milou}, the normalisation of the inelastic contribution is directly related to the exponential $t$-slope parameter.
The measured event yield corresponds to
an exponential $t$ distribution with a slope of $1.2$~GeV$^{-2}$
which is subsequently used in the simulation of inelastic DVCS events.
The corresponding contribution of proton dissociation
in the analysis event sample is found to be
$16 \pm 5$\%.
Other backgrounds from diffractive 
\om\ and \ph\ production with decay modes to final states including photons
are estimated to be negligible in the kinematic range of the analysis.
Contamination from processes with low multiplicity $\pi^0$ production was also investigated and found to be negligible.

The reconstruction method for the kinematic variables $Q^2$, $x$ and 
$W$ relies on the measured polar angles of the final 
state electron and photon (double angle method)~\cite{dvcsh1}.
The variable $t$ is approximated by the negative square of the 
transverse momentum of the outgoing proton.
The latter is computed from the vector sum of the transverse momenta 
of the final state photon $\vec P_{T_{\gamma}}$ and of the scattered 
electron $\vec P_{T_{e}}$:
$
  t \simeq - |\vec P_{T_{\gamma}} + \vec P_{T_{e}}|^2 \ .
$
The resolution of the $t$ reconstruction lies in the range $0.08$ to $0.22$ GeV$^2$.

\begin{figure}[!htbp]
\vspace*{2cm}
\begin{center}
 \includegraphics[totalheight=5.7cm]{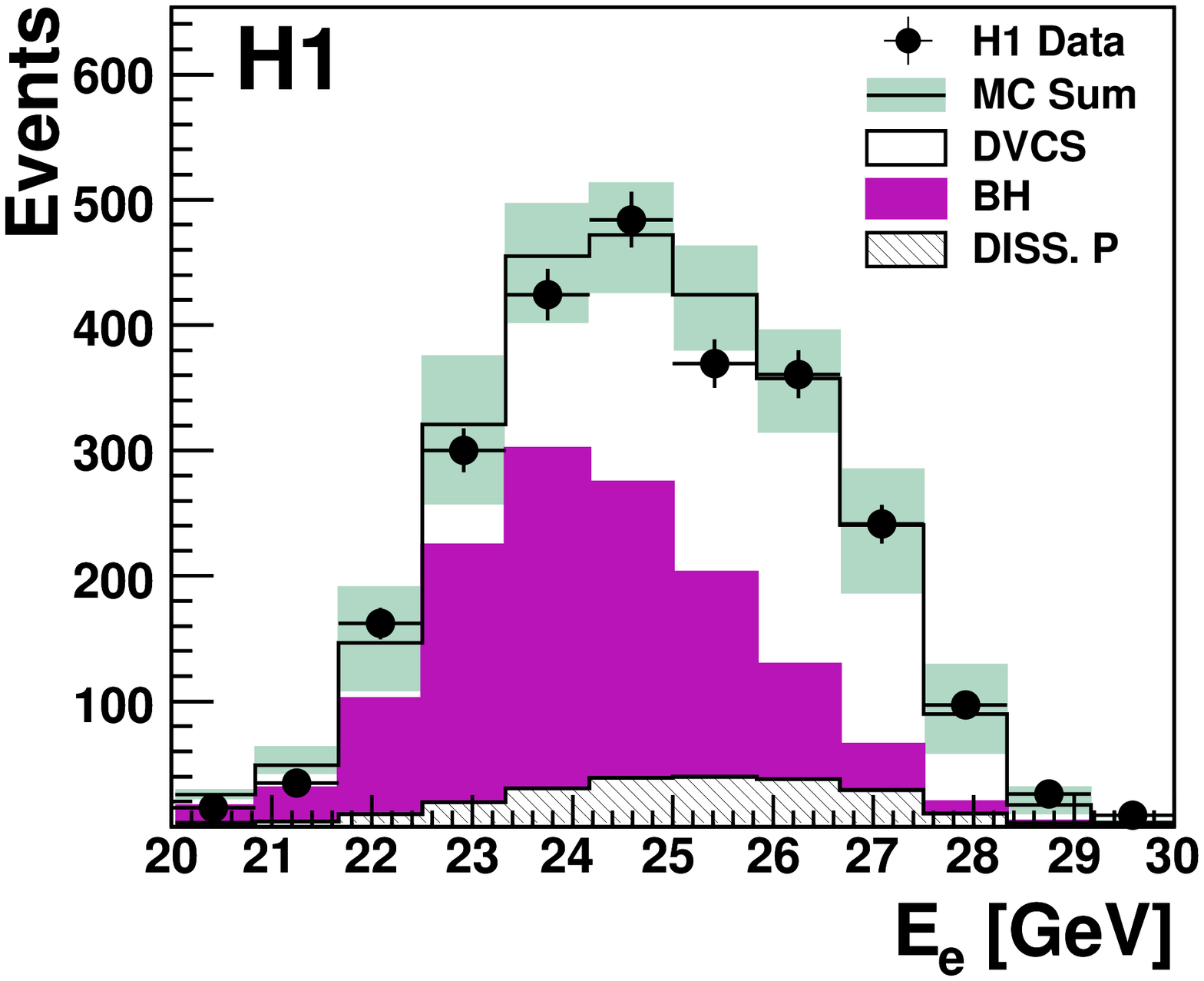}\put(-10,35){{(a)}}
 \includegraphics[totalheight=5.7cm]{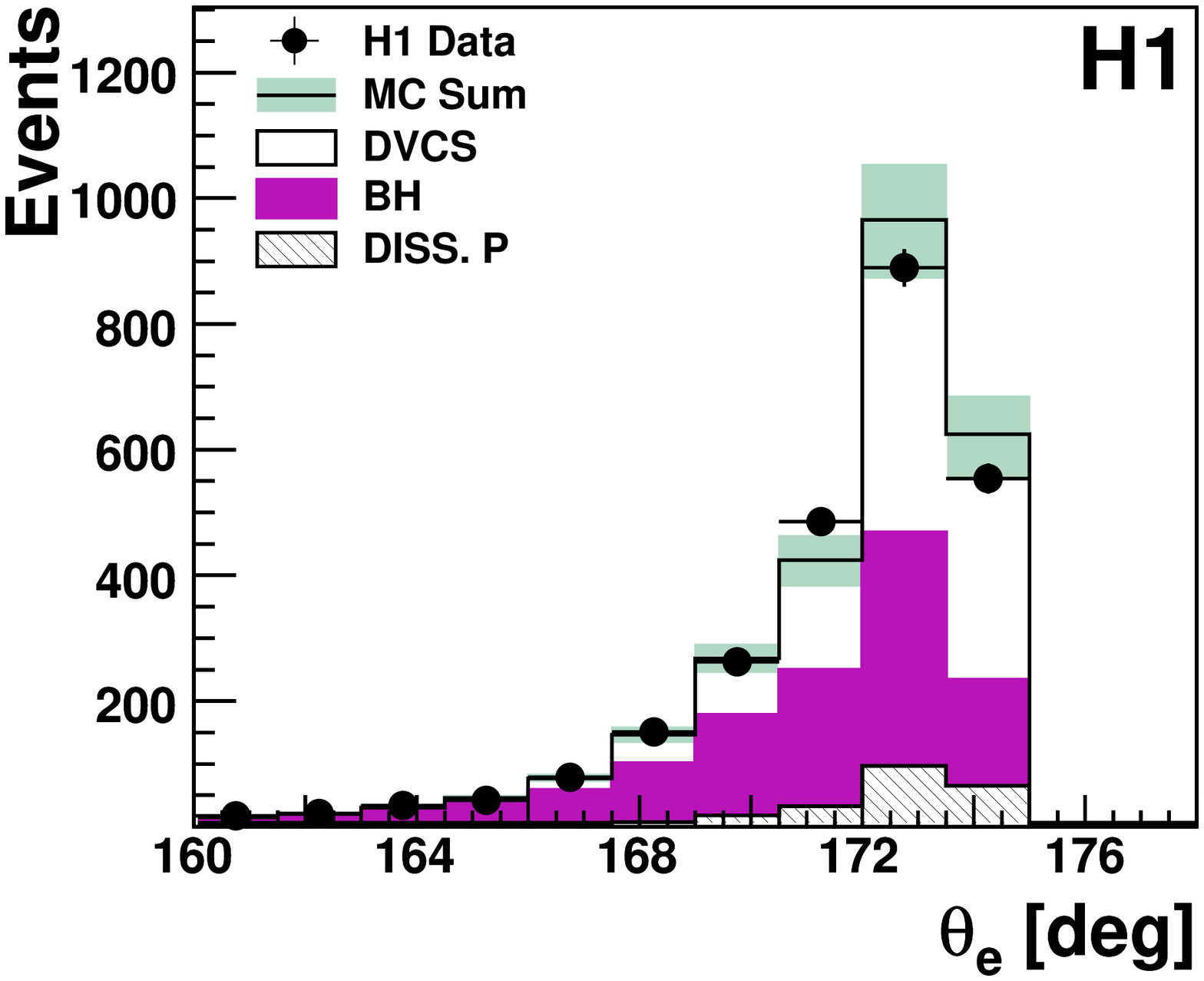}\put(-55,35){{(b)}}\\
 \includegraphics[totalheight=5.7cm]{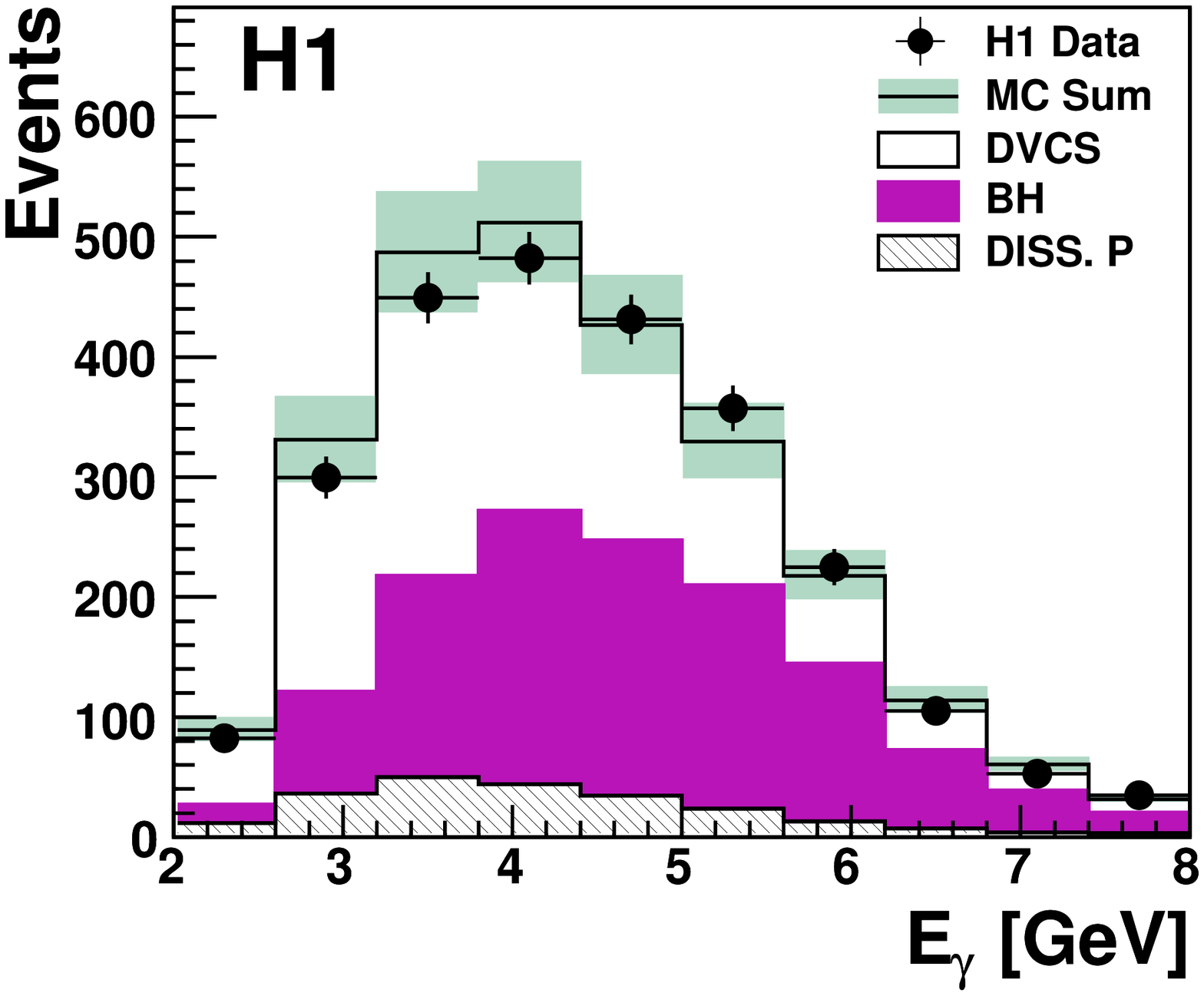}\put(-10,35){{(c)}}
 \includegraphics[totalheight=5.7cm]{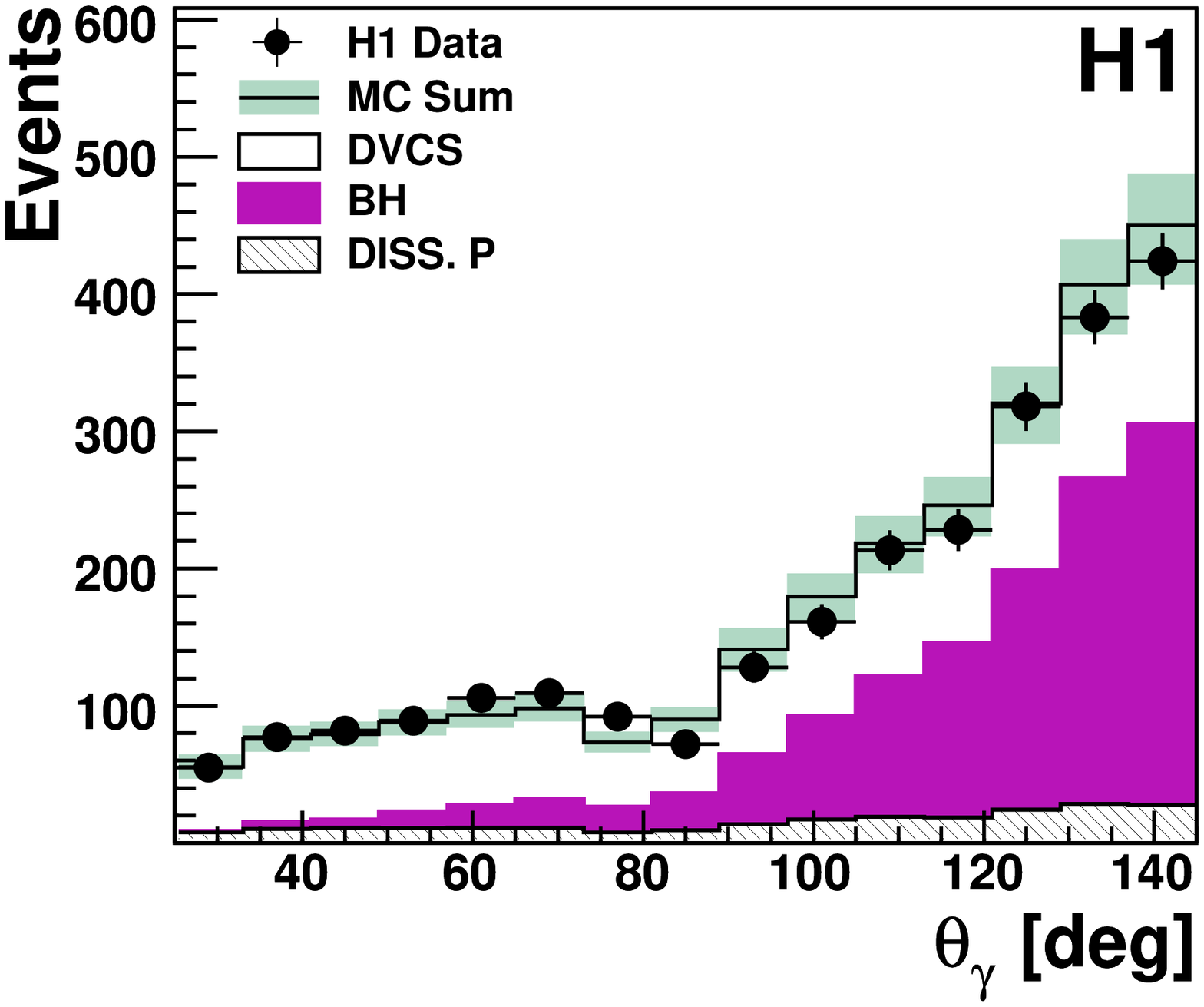}\put(-55,35){{(d)}}\\
 \includegraphics[totalheight=5.7cm]{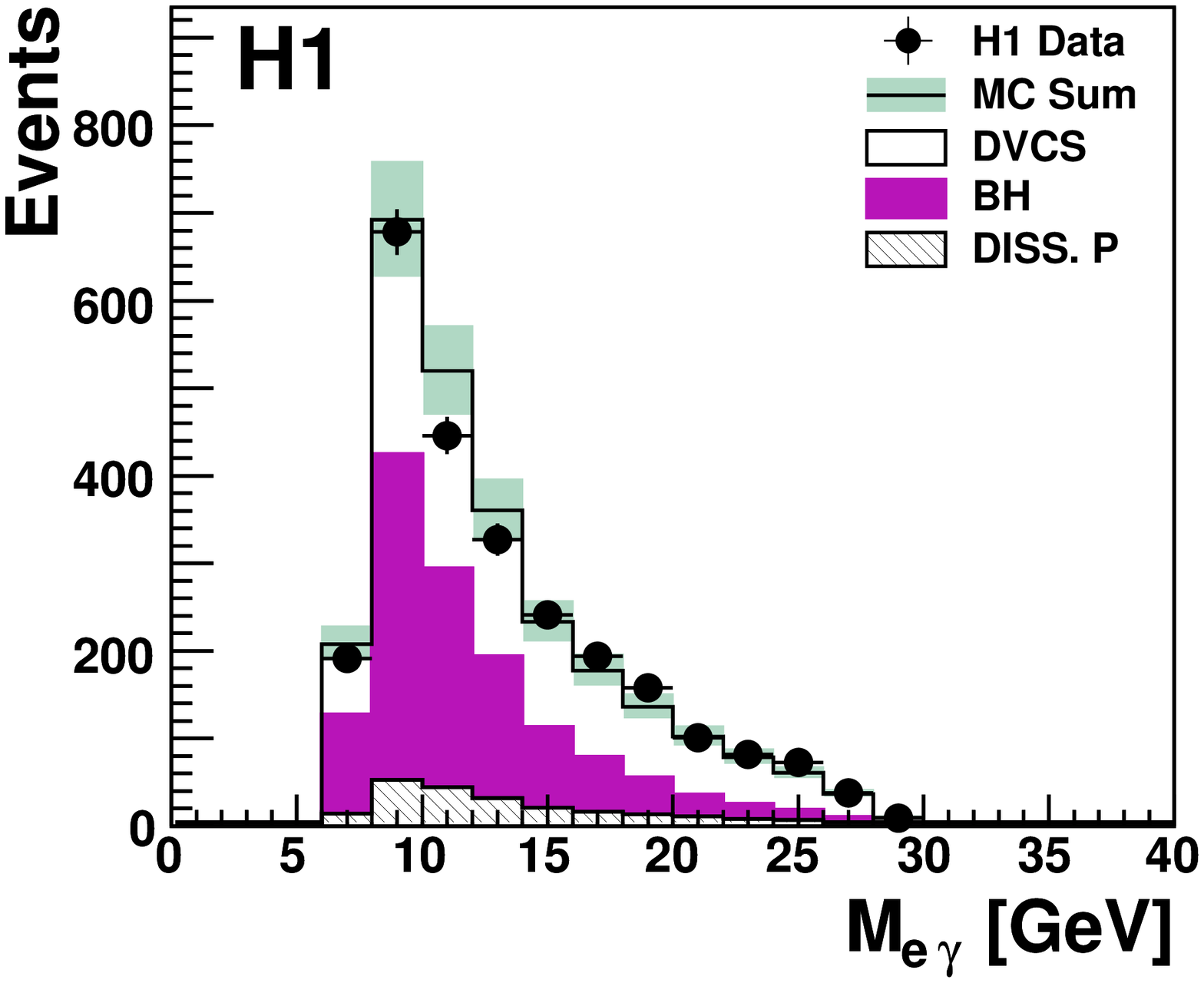}\put(-10,35){{(e)}}
 \includegraphics[totalheight=5.7cm]{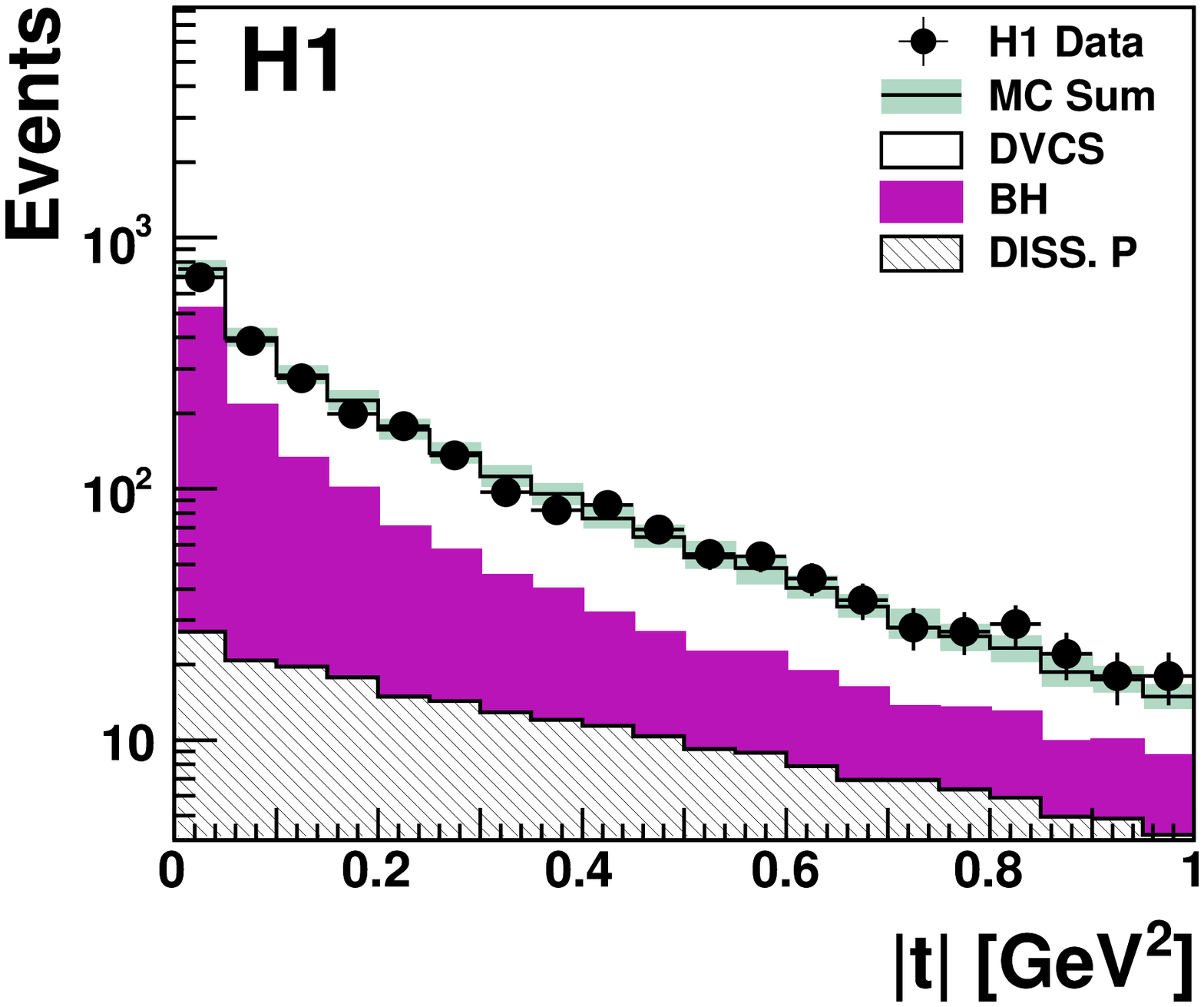}\put(-10,35){{(f)}}
\end{center}
\caption{\label{figcomp} 
Distributions of the energy (a) and polar angle (b) of the scattered
electron,  
the energy (c) and polar angle (d) of the photon, 
the electron-photon invariant mass (e) and the proton four momentum transfer squared $|t|$ (f).
The data are compared with Monte Carlo expectations for 
elastic DVCS, elastic and inelastic BH and inelastic DVCS (labelled DISS. p).
All Monte Carlo simulations are normalised according
to the luminosity of the data.
The open histogram shows the total
prediction and the shaded band its
estimated uncertainty.
}
\end{figure}

Distributions of selected kinematic variables are presented in figure~\ref{figcomp} for the analysis sample.
The MC expectations of the different processes are also displayed. 
Each source is normalised  to the data luminosity.
A good description of the shape and normalisation of the measured distributions is observed.

%=========================================================================
\section{Cross Section Determination and Systematic Uncertainties}
\label{sec-xsec}
%=========================================================================

The DVCS and BH contributions dominate in the analysis phase space.
In addition,
an interference term contributes to the cross section
due to the identical final states of both processes. 
In the leading twist approximation, the main contribution resulting from the 
interference of the BH and DVCS processes is proportional to the cosine 
of the azimuthal angle of the photon\footnote{The azimuthal angle of the photon is 
defined in the proton rest frame as the angle between the plane formed by the incoming and 
scattered electron and that formed by the virtual photon and the scattered proton.} \cite{bernard,phibel}.
Since the present measurement is integrated over this angle, 
the contribution of the interference term is 
estimated to be small (below $1$\%).
The DVCS cross section, $\gamma^\ast p \to \gamma p$, is evaluated in each bin $i$ with the bin centre values $Q^2_i,W_i,t_i$, from the total number $N^{\rm{obs}}_i$ of data events in the
analysis sample using the expression
\begin{equation}
\sigma_{DVCS}(Q^2_i,W_i,t_i)=
\frac{(N_i^{\rm{obs}}-N_i^{\rm{BH}} - N_i^{\rm{p-diss}})}{N_i^{\rm{DVCS}}}\cdot 
\sigma^{\rm{th}}_{DVCS}(Q^2_i,W_i,t_i).
\label{eq-ffs}
\end{equation}
The other numbers in this equation are calculated using the MC simulations described in section~\ref{simul}.
$N^{\rm{BH}}_i$ denotes the number of BH events (elastic and inelastic) reconstructed in bin $i$, $N^{\rm{p-diss}}_i$ the number of inelastic DVCS background events,
$N_i^{\rm{DVCS}}$ the number of DVCS events computed from the elastic DVCS MC and
$\sigma^{\rm{th}}_{DVCS}$ is the theoretical DVCS cross section used for the 
generation of DVCS MC events. 
The measured cross section is
thus directly corrected for detector inefficiencies and acceptances and is expressed at each bin centre value.

The mean value of the acceptance, defined as  the number of MC events reconstructed in a bin divided by the number of events generated in the same bin, is $45$\% over the whole kinematic range and reaches $78$\% for the highest $t$ bin. 
The  systematic errors of the measured DVCS cross section
are determined by repeating the analysis after applying to the MC appropriate variations
for each systematic source.
The main contribution comes from  the
acceptance correction factors calculated by varying the $t$-slope parameter set in the elastic DVCS MC by $\pm 8$\%. 
The uncertainty on the number of elastic DVCS events lost by the application of the FMD veto is modelled by a $4$\% variation of the FMD efficiency.
Both error sources together result in an error of $10$\% on the measured elastic DVCS cross section.
% The corresponding error on the cross section is of the order of $8$\%.
%
The uncertainty related to the inelastic DVCS background is estimated from the variation of its $t$-slope parameter by $25$\% around the nominal value of $b=1.2$~GeV$^{-2}$. The resulting error on the cross section amounts to $5$\% on average and reaches $15$\% at high $t$.
The uncertainties related to trigger efficiency, photon identification efficiency, 
radiative corrections and the subtraction of BH background and
luminosity measurement are each in the range of $2$ to $4$\%. 
The total systematic uncertainty of the cross section amounts to about $15$\% and is dominated by correlated errors.

\section{Results and Interpretations}

\subsection{Cross Sections and $t$-dependence}
\label{sec:results}
\begin{figure}[!htbp]
\begin{center}
 \includegraphics[totalheight=6cm]{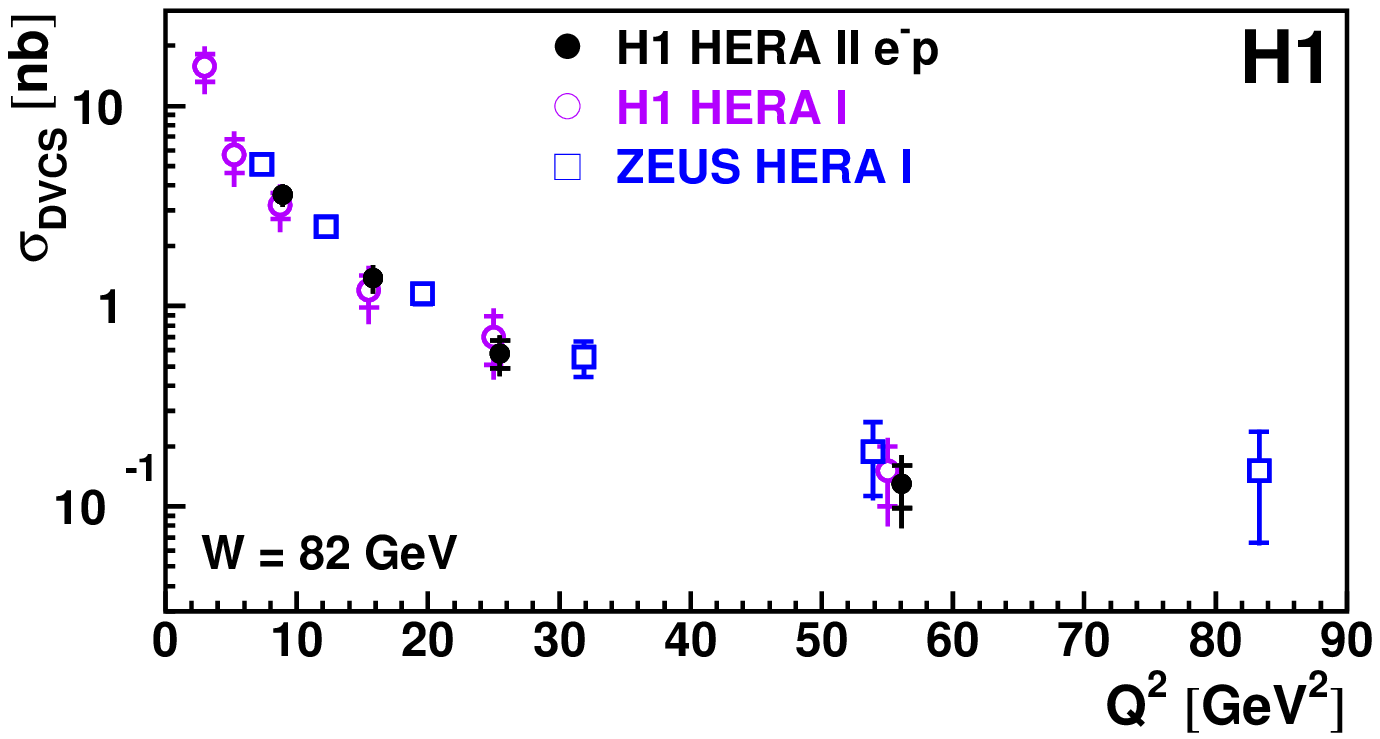}\put(-10,48){{(a)}}\\
 \includegraphics[totalheight=6cm]{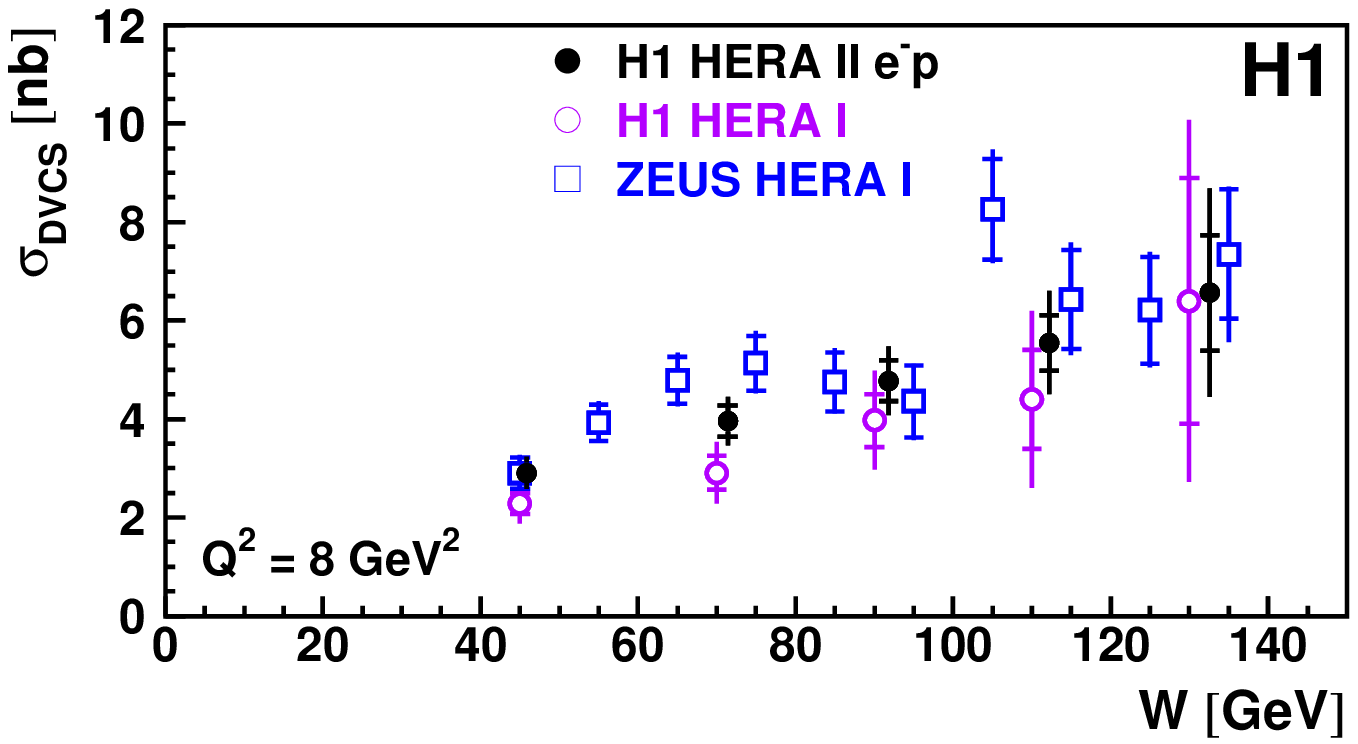}\put(-10,48){{(b)}}
\end{center}
\caption{\label{fig1d} 
  The DVCS cross section  as a function of
  $Q^2$ at  $W=82$~GeV (a) and as a function of
$W$ at $Q^2=8$~GeV$^2$ (b).
The results from the previous H1 and ZEUS publications \cite{dvcsh1,dvcszeus} based on HERA I data
are also displayed.
The inner error bars represent the statistical errors, 
the outer error bars the statistical and systematic errors added in quadrature.
}
\end{figure}

The complete DVCS sample is used to extract the $W$ dependence of the  DVCS  cross section expressed at $Q^2=8$~GeV$^2$ as well as the $Q^2$ dependence at $W=82$~GeV. 
The results are displayed in figure~\ref{fig1d} and are in agreement
within errors with the previous measurements~\cite{dvcsh1,dvcszeus}. 
The steep rise of the cross section with $W$ 
is an indication of the presence of a hard underlying process~\cite{ivanov}.
The corresponding cross section measurements are shown in table~\ref{sig1d}.

\begin{figure}[!htbp]
\begin{center}
 \includegraphics[totalheight=6cm]{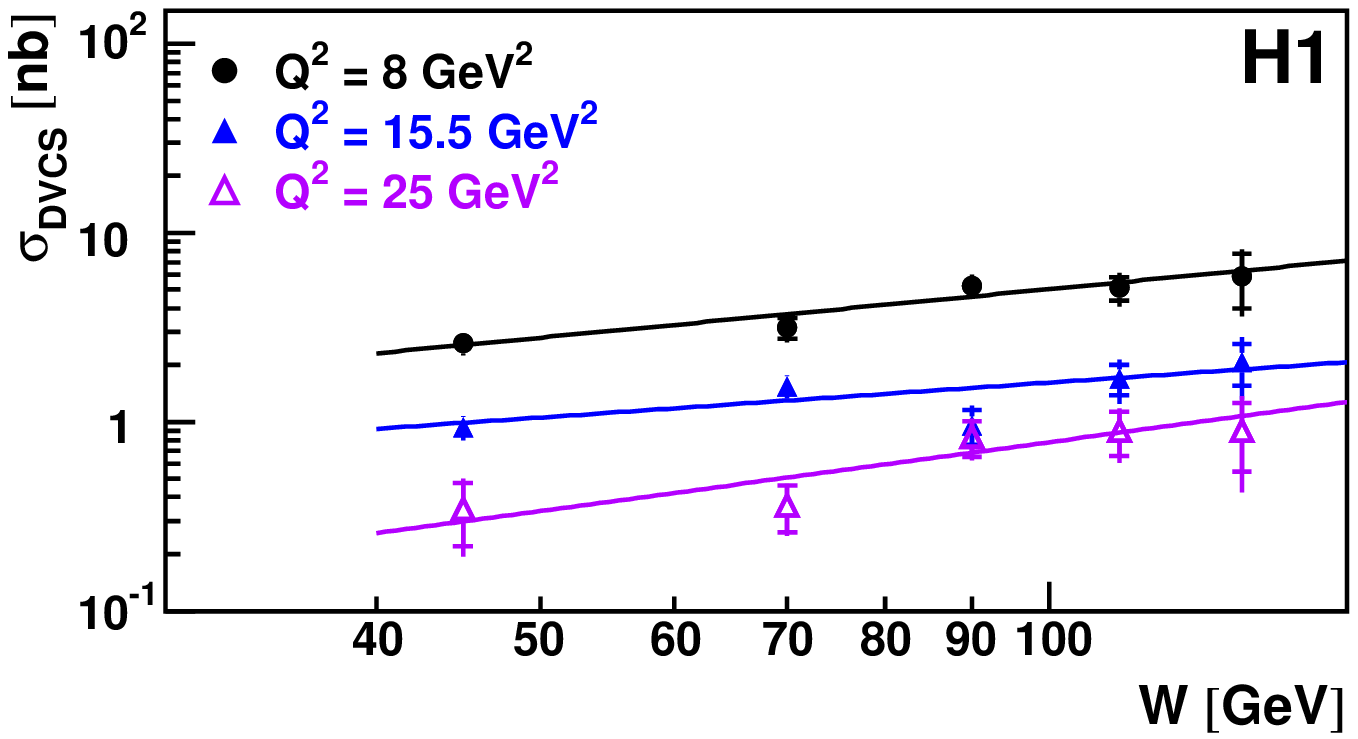}\put(-10,48){{(a)}}\\
 \includegraphics[totalheight=6cm]{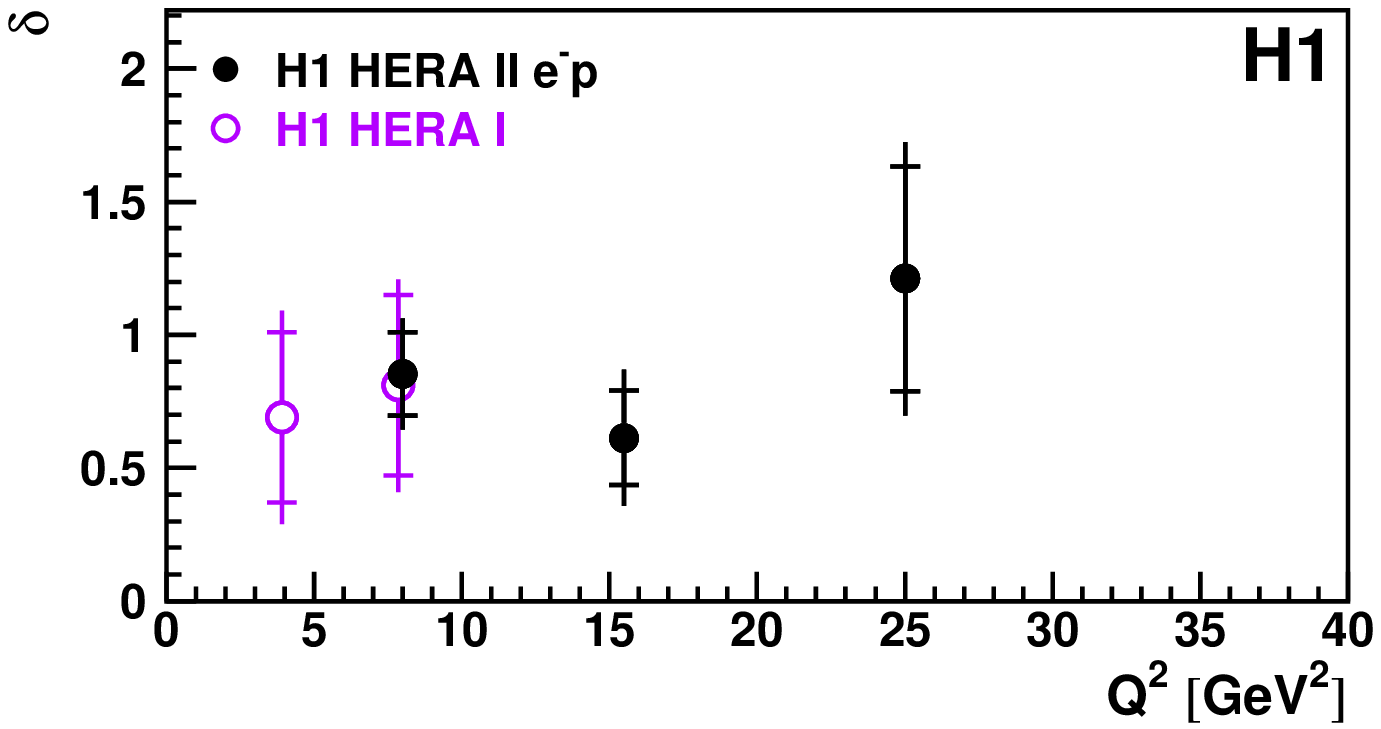}\put(-10,48){{(b)}}
\end{center}
\caption{\label{fig2d} 
  The DVCS cross section as a function of
  $W$ at three values of $Q^2$ (a). The solid lines represent the results
of fits of the form $W^\delta$. The fitted values of $\delta(Q^2)$
are shown in (b). 
The inner error bars represent the statistical errors, 
the outer error bars the statistical and systematic errors added in quadrature.
}
\end{figure}

Next, the $W$ dependence of the DVCS  cross section is determined for three separate ranges of $Q^2$ and shown in figure~\ref{fig2d}(a). 
The corresponding cross section measurements are given in table~\ref{sig2d}.
A fit of the form $W^\delta$ is performed to the cross section in each $Q^2$ range.
Figure~\ref{fig2d}(b) presents the $\delta$ values obtained as a function of $Q^2$. It is observed that $\delta$ is independent
of $Q^2$ within the errors. 
Using the complete analysis sample, the value of $\delta$ expressed at 
 $Q^2 = 8\; \mbox{GeV}^2$ is found to be
$0.74 \, \pm \, 0.11 \, \pm \, 0.16$, where the first error is statistical and the second systematic.

\begin{figure}[!htbp]
\begin{center}
\includegraphics[width=9.5cm]{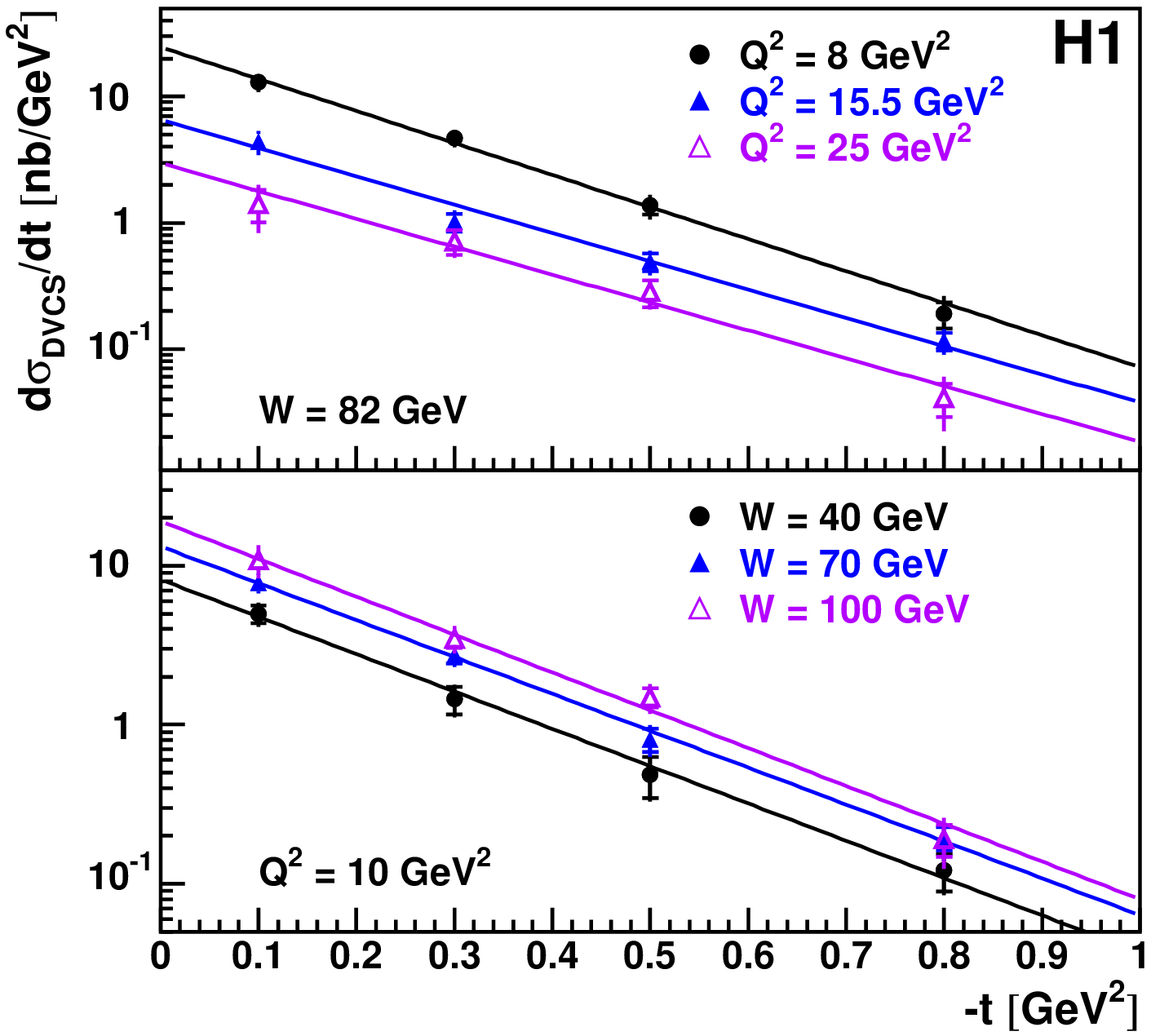}\put(-8,70){{(a)}}\put(-8,32){{(b)}}\\
 \includegraphics[width=9.5cm]{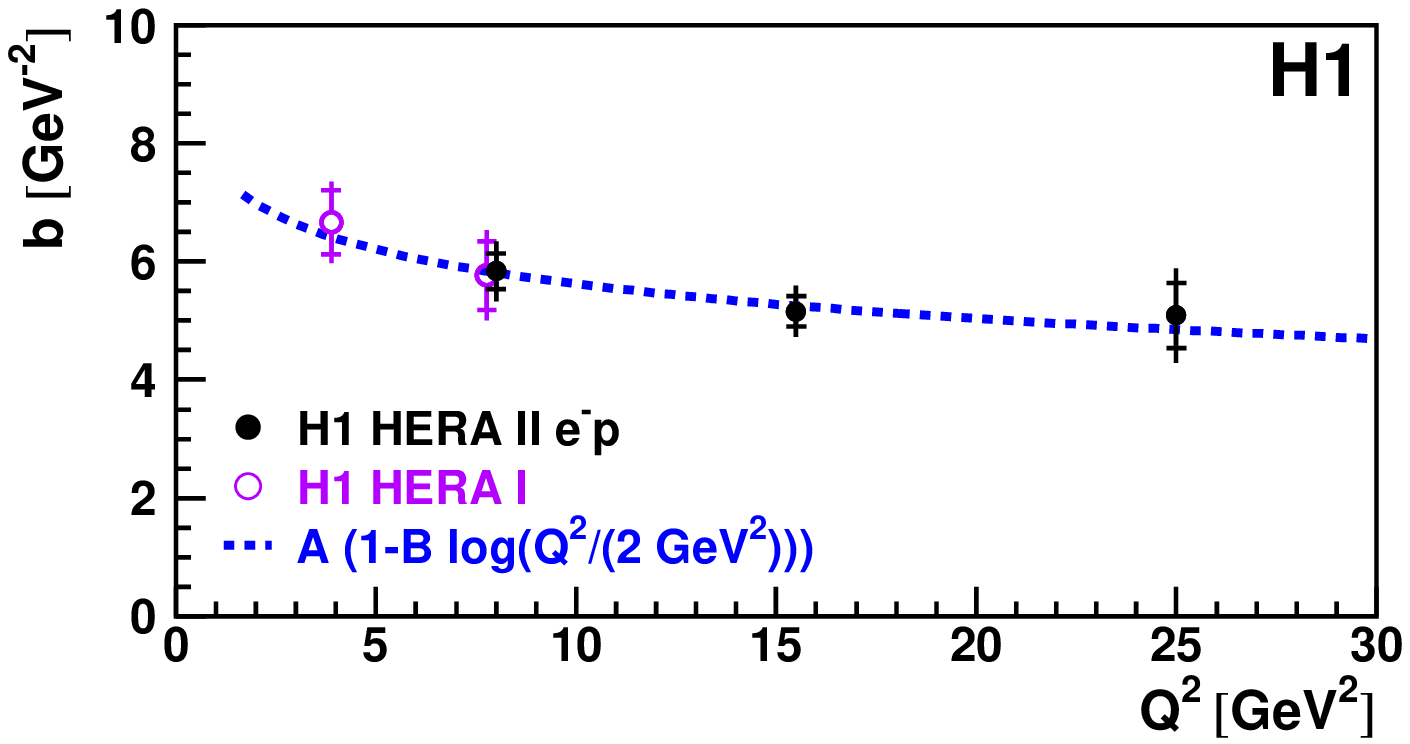}\put(-8,34){{(c)}}\\
 \includegraphics[width=9.5cm]{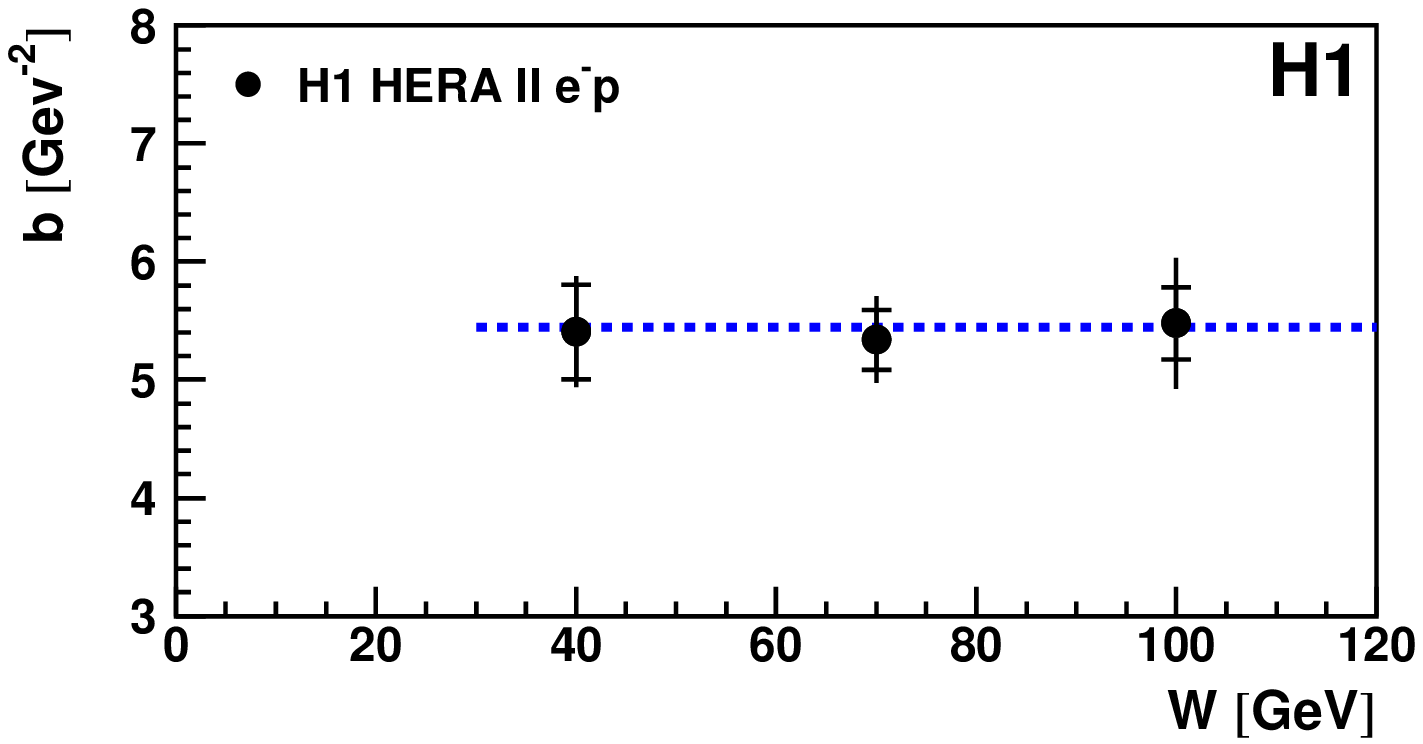}\put(-8,34){{(d)}}
\end{center}
\vspace*{-0.5cm}
\caption{\label{figb} 
The DVCS cross section, differential in $t$, 
for three values of $Q^2$ expressed at \mbox{$W=82$~GeV (a)} and 
for three values of $W$ at $Q^2=10$~GeV$^2$ (b).
The solid lines in (a) and (b) represent the results of fits 
of the form $e^{-b|t|}$.
The fitted $t$-slope parameters $b(Q^2)$ are shown in (c) together with the
$t$-slope parameters from the previous H1 publication \cite{dvcsh1}.
The dashed curve in (c) represents the result of a fit to the $b(Q^2)$ values using a phenomenological function as described in the text.
In (d) the fitted $t$-slope parameters  $b(W)$ are shown. 
The dashed line in (d) corresponds to the 
average value $b=5.45 $~GeV$^{-2}$, obtained from
a fit to the complete data sample of the present measurement.
The inner error bars represent the statistical errors
and the outer error bars the statistical and systematic errors added in quadrature.
}
\end{figure}

The differential cross section as a function of $t$ is displayed in figures~\ref{figb}(a) and (b)
 for three values of $Q^2$ and $W$, respectively.
Fits of the form $d\sigma/dt \sim e^{-b|t|}$ are performed taking into account the statistical and correlated systematic errors; they describe the data well.
The derived $t$-slope parameters $b(Q^2)$ and $b(W)$ are displayed in figures~\ref{figb}(c) and (d), respectively.
The cross section values and the results for $b$ in each $Q^2$ and $W$ bin are given in table~\ref{sigtq}. 
This analysis extends the study of the evolution of $b$ with $Q^2$ to larger values than 
in the previous H1 measurement~\cite{dvcsh1}. 
This  $Q^2$ dependence can be parametrised~\cite{freund2} as
\begin{equation}
b(Q^2)=A \left( 1-B \log(Q^2/(2 \; \mbox{GeV}^2)) \right).
\label{bq2}
\end{equation}
Fitting this function to the measured $b$ values of the present data and to the value obtained at $Q^2=4$~GeV$^2$ in the previous H1 publication~\cite{dvcsh1} yields $A=6.98 \pm 0.54$ GeV$^2$ and $B=0.12 \pm 0.03$.
The systematic errors and their point to point correlations were taken into account in the fit, resulting in a correlation coefficient between
$A$ and $B$ of $\rho_{AB} = 0.92$.
As shown in figure~\ref{figb}(c) the fit function provides a good description of the measured $b$ values over the whole $Q^2$ range.
The values of $b$ as a function of $W$ are measured for the first time and shown in figure~\ref{figb}(d). No significant variation of $b$ with $W$ is observed.

Using the complete analysis sample, the value of $b$ expressed at $Q^2 = 8\; \mbox{GeV}^2$ is found to be $5.45 \, \pm \, 0.19 \, \pm \, 0.34$~GeV$^{-2}$,  where the first error is statistical and the second systematic.
Following~\cite{bukpaper,strikpaper}, this $t$-slope  value can be converted to an average impact parameter of $\sqrt{<r_T^2>} = 0.65 \pm 0.02$~fm. 
It corresponds to the transverse extension
of partons, dominated by sea quarks and gluons for an average value $x =1.2 \ 10^{-3}$, 
in the plane perpendicular to the direction of motion of the proton. 
This value is related to the size of the core of the proton
with no account of the peripheral soft structure.
%

%=========================================================================
\subsection{QCD Interpretation in Terms of GPDs}
%=========================================================================

\begin{figure}[!htbp]
\begin{center}
 \includegraphics[totalheight=11cm]{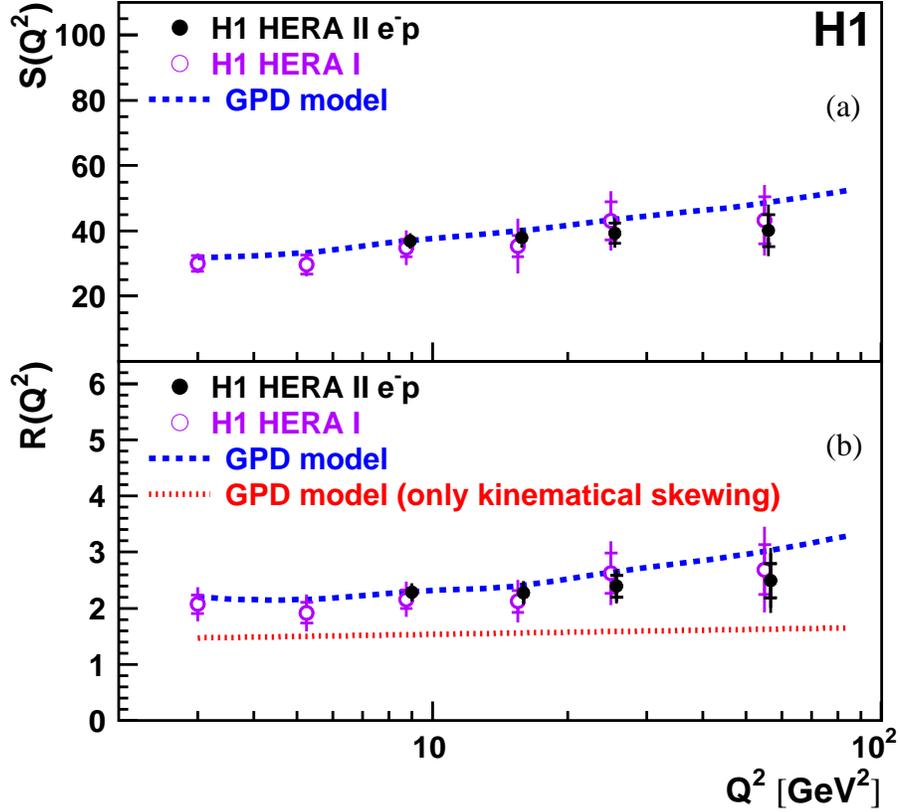}\put(-10,93){{(a)}}\put(-10,48){{(b)}}
\end{center}
\vspace*{-0.6cm}
\caption{\label{figs} 
The observables $S$ and $R$ (see text), shown as a function of $Q^2$ in (a) and (b), respectively.
The results from the previous H1 publication \cite{dvcsh1} based on HERA I data
are also displayed.
The inner error bars represent the statistical errors, 
the outer error bars the statistical and systematic errors 
added in quadrature.
The dashed curves show the predictions
of the GPD model \cite{freund2,cteq}.
In (b), the dotted curve shows the prediction 
of a GPD model based on an approximation 
where only the kinematical part of the skewing effects are taken into account (see text).
}
\end{figure}

%
The determination of $b(Q^2)$ described above can be used to study the $Q^2$ evolution of the GPDs.
The DVCS cross section integrated over the momentum transfer $t$ can be written~\cite{ffspaper} as
\begin{equation}
 \sigma_{DVCS} (Q^2,W)
  \equiv   \frac{\left[ \,{\cal I}m {A}(\gamma^*p \to \gamma p)_{t=0}(Q^2,W)\right]^2 (1+\rho^2)}{16\pi\,b(Q^2,W)} \ ,
\label{test1}
\end{equation}
where ${\cal I}m {A}(\gamma^*p \to \gamma p)_{t=0}(Q^2,W)$ is the imaginary part of the $\gamma^*p \to \gamma p$ scattering amplitude at $t=0$ and $\rho^2$ is a small
correction due to the real part of the amplitude.
In the following, $\rho$ is determined from dispersion relations~\cite{favart} to be $\rho=\tan(\frac{\pi}{2} \omega(Q^2))$. 
The coefficient $\omega(Q^2)$  
describes the power governing
the $W$ dependence of DVCS at a given $Q^2$.
It is taken from the corresponding power of
the rapid rise of the proton structure function
$F_2$ at low $x$ ($F_2 \sim x^{-\omega}$)
~\cite{h1f2bis}, assuming that it is sufficiently close
to the one in DVCS.
In the GPD formalism, 
the amplitude ${A}(\gamma^*p \to \gamma p)_{t=0}$ is directly proportional to the GPDs.
As shown in the previous section, the $Q^2$ dependence of the $t$-slope $b$ is non-negligible.
Therefore, the $Q^2$ evolution of the GPDs themselves is accessed by removing this variation of $b(Q^2)$.
For this purpose, the dimensionless observable $S$ is defined as
\begin{equation}
S = \sqrt{ \frac{{\sigma_{DVCS} \ Q^4 \ b(Q^2)}} {{(1+\rho^2)}}} \ .
\end{equation}
Using the parametrisation~(\ref{bq2}) for $b(Q^2)$, $S$ is then calculated for each $Q^2$ bin from the cross section measurements of this analysis (table~\ref{sig1d}) and from those of the previous H1 publication~\cite{dvcsh1}.
The uncertainties on the parameters $A$ and $B$ of~(\ref{bq2}) are directly propagated to determine the error on $b(Q^2)$ at any given $Q^2$ value. 
The results for $S$ are presented in figure~\ref{figs}(a) together with the prediction of a GPD model~\cite{freund2}, based on the PDFs parametrisation given in~\cite{cteq}. 
It is observed that the pQCD skewed evolution equations~\cite{diehlpaper,mueller,guzey} provide a reasonable description of the measured weak rise of $S$ with $Q^2$. 

The magnitude of the skewing effects present in the DVCS process can be extracted by constructing the ratio of the imaginary parts of the DVCS and DIS amplitudes.
At leading order in $\alpha_s$, this ratio 
$
R \equiv 
{{\cal I}m \,{{ A}}\,(\gamma^* p \to \gamma  p)_{t=0}} /
{{\cal I}m\, {{  A}}\,(\gamma^* p \to \gamma^*  p)_{t=0}} \
$
is equal to
the ratio of the GPDs to the PDFs.
The virtual photon is assumed to be mainly transversely polarised 
in the case of the DVCS process due to the real photon 
in the final state and therefore 
has to be taken as transversely polarised in the DIS amplitude too. 
The expression for $R$ as a function of the measured observables can be written as
\begin{equation}
 R =\frac
 {4\,\sqrt{\pi \ \sigma_{DVCS}  \ b(Q^2)}}
 {\sigma_T(\gamma^* \,p\rightarrow X)\, \sqrt{(1+\rho^2)}} \,
= \frac
 {\sqrt{\sigma_{DVCS} \ Q^4 \ b(Q^2)}}
 {\sqrt{\pi^3} \,\alpha_{EM} F_T(x,Q^2)\, \sqrt{(1+\rho^2)}} \   , 
\label{R_def_ap} 
\end{equation}
using the relation $\sigma_T(\gamma^* \,p\rightarrow X) = 4\pi^2 \alpha_{EM} F_T(x,Q^2)/ Q^2$
with $\alpha_{EM}=1/137$.
$R$ is evaluated taking
 $F_T = F_2 - F_L$ from the QCD analysis presented in~\cite{h1f2} and using the parametrisation~(\ref{bq2}) for $b(Q^2)$.
The measured values of the ratio $R$ for each $Q^2$ bin are shown in figure~\ref{figs}(b) and compared with the calculation based on the GPD model proposed in~\cite{freund2}.
%
%The error on the GPD curve coming from the use of a different $F_T$ parametrisation in the GPD model and in equation~\ref{R_def_ap} is below $3$\%.
%
The typical values of $R$ are around $2$, whereas in a model without skewing $R$ would be equal to unity. 
Therefore, the present measurement confirms the large effect of skewing.
In GPD models, two different effects contribute to skewing~\cite{diehlpaper,mueller,guzey}: the kinematics of the DVCS process and the $Q^2$ evolution of the GPDs. 
The data are compared to a model which takes only the former effect into account.
The result of this incomplete model is represented by a dotted line in figure~\ref{figs}(b). 
The present measurements show that
such an approximation is not sufficient to reproduce the total skewing effects observed in the data.

%=========================================================================
\subsection{Geometric Scaling}
\label{geom}
%=========================================================================
%

\begin{figure}[!htbp]
\vspace*{-1cm}
\begin{center}
 \includegraphics[totalheight=10.2cm]{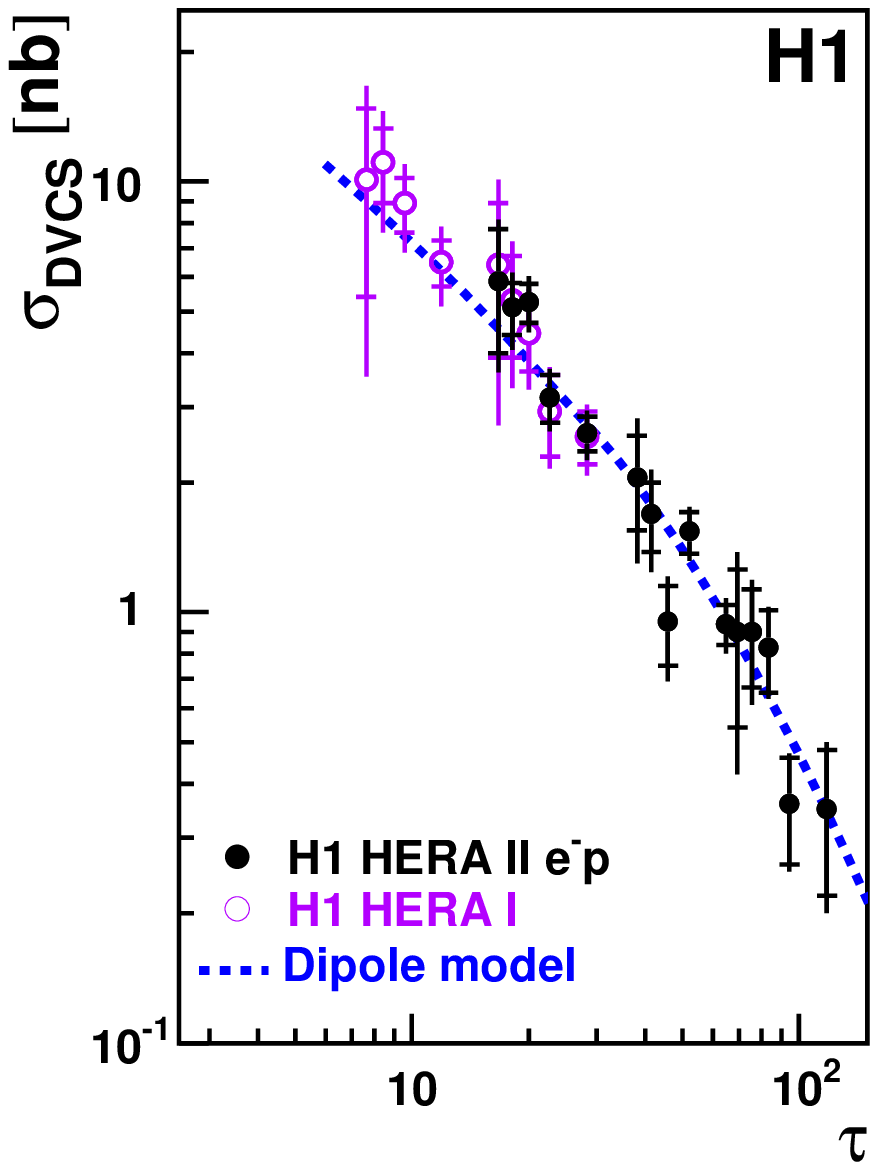}\put(-9,75){{(a)}}
 \includegraphics[totalheight=10.2cm]{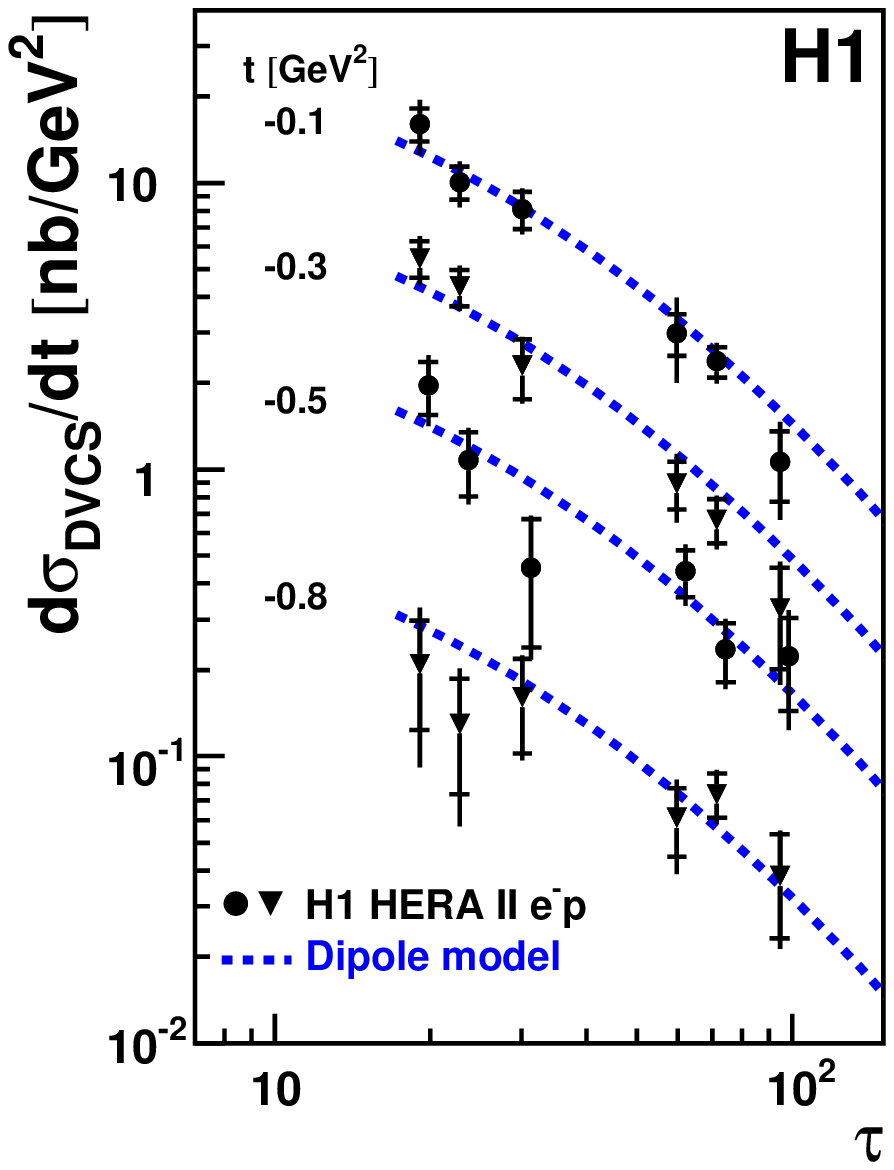}\put(-9,75){{(b)}}
\end{center}
\vspace*{-1.cm}
\caption{\label{figgs0} 
DVCS cross section measurements as a function of 
the scaling variable
$\tau={Q^2}/{Q_s^2(x)}$. 
In (a) the results are shown
for the full $t$ range  $|t| <$ 1 GeV$^2$
and in (b) at four values of $t$.
The cross section measurements from the previous H1 publication \cite{dvcsh1}
are also shown in (a).
The inner error bars indicate the statistical errors, 
the outer error bars the statistical and systematic errors added in quadrature.
The dashed curves represent the predictions of the 
dipole model \cite{lolo,iim2}.
}
\end{figure}

As discussed in section~\ref{sec:intro}, the dipole model represents another possible theoretical approach to describe the DVCS reaction.
It is therefore interesting to test if the present DVCS measurements obey the geometric scaling laws predicted by such models.
In the following study parameters of the dipole model are taken from an analysis of the total DIS cross section~\cite{iim2,robilast}.
The saturation scale $Q_s(x)=Q_0 ({{x_0}/{x}})^{-\lambda/2}$ is evaluated using $Q_0= 1$ GeV, $\lambda=0.25$ and $x_0=2.7\ 10^{-5}$.
The DVCS cross section measurements listed in table~\ref{sig2d} and those from the previous H1 publication~\cite{dvcsh1} which are measured at different $Q^2$ and $x=Q^2/W^2$ values can be represented as a function of the single variable $\tau$ (see equation~(\ref{eq:tau})).
The result is shown in figure~\ref{figgs0}(a). 
All of the cross section measurements appear to be well aligned on a single curve as a function of $\tau$. 
Therefore the DVCS data are compatible with the geometric scaling law.
The dipole model of~\cite{lolo,iim2} is also represented in figure~\ref{figgs0}(a) and
gives a good description of the cross section measurements over the complete range of $\tau$.

The dependence of the DVCS cross section on $\tau$ is also studied at
four different values of $t$. 
For this purpose, the cross section is measured differentially
in $t$ for three values of $W$ and two ranges of $Q^2$
($6.5 < Q^2 < 11$~GeV$^2$ and $11 < Q^2 < 80$~GeV$^2$), as listed in table~\ref{tabgs1}. 
Keeping the same parameters $Q_0$, $\lambda$ and $x_0$ as previously defined, each value of the differential cross section is again represented as a function of $\tau$.
The results are shown in figure~\ref{figgs0}(b), together with  the 
predictions of the dipole model~\cite{lolo,iim2}.
For these predictions, the $t$-dependence is factorised out as $e^{-b|t|}$, where the global $t$-slope parameter $b$ measured in section~\ref{sec:results} is used.
A reasonable description of the DVCS cross section values in the four $t$ bins is observed, with the same saturation scale $Q_s(x)$ used in all cases.

%=========================================================================
\section{Conclusion}
%=========================================================================

The cross section for deeply virtual Compton scattering $\gamma^\ast p \rightarrow \gamma p$
has been measured with the H1 detector at HERA.
The analysis uses the $e^{-}p$\/ data recorded in $2005$ and $2006$
corresponding to a luminosity of $145$ pb$^{-1}$, four times larger than in the previous H1 publication~\cite{dvcsh1}.
The measurement is performed in the 
kinematic range $6.5<Q^2<80$~GeV$^2$,
$30~<~W~<~140$~GeV and $|t| <$~1~GeV$^2$.

The $W$ dependence of the cross section is well described by a function
$W^{\delta}$.
No significant variation of the exponent $\delta$ as a function of $Q^2$ is observed.
For the total sample a value $\delta = 0.74 \, \pm \, 0.11 \, \pm \, 0.16$ is determined.
The steep rise of the cross section with $W$ indicates a hard underlying process.
The $t$-dependence of the cross section is well described by the form $e^{-b|t|}$
with an average slope of 
$b=5.45 \pm 0.19 \pm 0.34$~GeV$^{-2}$. 
This value corresponds to a transverse extension
of sea quarks and gluons in the proton of $\sqrt{<r_T^2>} = 0.65 \pm 0.02$~fm.
The $t$-slopes are determined for the first time differentially in $W$ with no
significant dependence observed.
The study of the $Q^2$ dependence of $b$ is extended to significantly
larger $Q^2$ values compared to previous measurements.
The slopes found in the present analysis and in the previous H1 publication
are in agreement with a slow decrease of $b$ as a function of $Q^2$.

The measurement of $b(Q^2)$ obtained in the present analysis is used to constrain the normalisation and $Q^2$ dependence of theoretical predictions based on GPDs. 
It is found that a GPD model reproduces well both the DVCS amplitude and its   
weak rise with $Q^2$.
The skewing effects have been investigated and are found to be large, as expected in GPD models.
Another approach based on a dipole model including saturation effects
predicts that the cross section can be approximated by a function
of the single variable, $\tau=Q^2/Q^2_s(x)$ where $Q_s(x)$ is the saturation scale. 
The present measurement of the DVCS cross section is found to be compatible 
with such a geometric scaling using the same parameters as derived from inclusive DIS.
For the first time, this scaling property is observed for different values of $t$.

%=========================================================================
\section*{Acknowledgements}
%=========================================================================

We are grateful to the HERA machine group whose outstanding
efforts have made this experiment possible. 
We thank the engineers and technicians for their work in constructing 
and maintaining the H1 detector, our funding agencies for financial 
support, the DESY technical staff for continual assistance and the 
DESY directorate for the hospitality which they extend to the non DESY 
members of the collaboration.
We would like to thank Markus Diehl for helpful discussions.

%=========================================================================

%=========================================================================

%=========================================================================
\vfill
\newpage

\begin{table}[htbp]
\centering
\begin{tabular}{|c|lll|l|c|lll|}
\cline{1-4} \cline{6-9} %\\[-10pt]
    & 
   \multicolumn{3}{c|}{} &
\hspace{0.5 cm} &   & 
  \multicolumn{3}{c|}{} \\[-10pt]
  $Q^2$ $\left[{\rm GeV}^2 \right]$  & 
   \multicolumn{3}{c|}{$\sigma_{DVCS}$ $\left[{\rm nb}\right]$} &
\hspace{0.5 cm} & $W$ $\left[{\rm GeV}\right]$  & 
  \multicolumn{3}{c|}{$\sigma_{DVCS}$ $\left[{\rm nb}\right]$} \\[1.5pt]
\cline{1-4} \cline{6-9}%\\[-10.0pt]
    & 
   \multicolumn{3}{c|}{} &
\hspace{0.5 cm} &   & 
  \multicolumn{3}{c|}{} \\[-12pt]
 $8.75$  & $3.59$  &$\pm$ $0.21 $&$\pm$ $0.41 $ & &  $ 45$ & $2.91$  &$\pm$ $0.20$&$\pm$ $0.25 $ \\
 $15.5$  & $1.38$  &$\pm$ $0.10 $&$\pm$ $0.21 $ & &  $ 70$ & $3.96$  &$\pm$ $0.32$&$\pm$ $0.37 $ \\
 $25$    & $0.58$  &$\pm$ $0.09$ &$\pm$ $0.09 $  & & $ 90$ & $4.78$  &$\pm$ $0.41$&$\pm$ $0.57 $ \\
 $55$    & $0.13$  &$\pm$ $0.03$ &$\pm$ $0.04 $  & & $110$ & $5.55$  &$\pm$ $0.57$&$\pm$ $0.88 $ \\
         &         &           &             & & $130$ & $6.56$  &$\pm$ $1.17$&$\pm$ $1.77 $ \\[1.5pt]
\cline{1-4} \cline{6-9}
\end{tabular}
\caption{ 
 The DVCS cross section $\gamma^\ast p \rightarrow \gamma p$, $\sigma_{DVCS}$,
 as a function of $Q^2$ for 
 $W=82\,{\rm GeV}$ and as a function of $W$ for $Q^2=8\,{\rm GeV}^2$ , both 
 for $ |t| < 1\,{\rm GeV}^2$.
 The first errors are statistical, the second systematic.}
\label{sig1d}
\end{table}

\begin{table}[!htbp]
\centering
\begin{tabular}{|c|lcc|lcc|lcc|}
 \cline{2-10}
 \multicolumn{1}{c|}{~} &\multicolumn{9}{c|}{~} \\ [-10pt]
 \multicolumn{1}{l|}{} & \multicolumn{9}{c|}{$\sigma_{DVCS}
    \; \; \left[{\rm nb}\right]$} \\ [3.0pt]
%  \cline{2-10}
%  \multicolumn{1}{c|}{~} &\multicolumn{3}{c|}{~}  
%    &\multicolumn{3}{c|}{~}  
%    &\multicolumn{3}{c|}{~} \\ [-10pt]
 \hline
 \multicolumn{1}{|c|}{~} &\multicolumn{3}{|c|}{~}  
   &\multicolumn{3}{c|}{~}  
   &\multicolumn{3}{c|}{~} \\ [-10pt]
 $W$ $\left[{\rm GeV}\right] $  &
  \multicolumn{3}{c|}{$Q^2 = 8 \;$GeV$^2$} &  
  \multicolumn{3}{c|}{$Q^2 = 15.5 \;$GeV$^2$} & 
  \multicolumn{3}{c|}{$Q^2 = 25 \;$GeV$^2$} \\ [3.0pt]
 \hline 
      $45$&       $2.60$&    $\pm$   $0.24$&  $\pm$     $0.24$&    $0.94$&   $\pm$    $0.10$&  $\pm$	$0.10$&	    $0.35$&   $\pm$    $0.13$&  $\pm$  $0.07$ \\
      $70$&       $3.15$&    $\pm$   $0.40$&  $\pm$     $0.33$&    $1.54$&   $\pm$    $0.17$&  $\pm$	$0.14$&	    $0.36$&   $\pm$    $0.10$&  $\pm$  $0.05$ \\
      $90$&       $5.25$&    $\pm$   $0.55$&  $\pm$     $0.55$&    $0.95$&   $\pm$    $0.20$&  $\pm$	$0.17$&	    $0.83$&   $\pm$    $0.18$&  $\pm$  $0.09$ \\
     $110$&       $5.11$&    $\pm$   $0.71$&  $\pm$     $0.76$&    $1.69$&   $\pm$    $0.31$&  $\pm$	$0.33$&	    $0.90$&   $\pm$    $0.23$&  $\pm$  $0.18$ \\
     $130$&       $5.88$&    $\pm$   $1.89$&  $\pm$     $1.26$&    $2.06$&   $\pm$    $0.51$&  $\pm$	$0.56$&	    $0.90$&   $\pm$    $0.36$&  $\pm$  $0.32$ \\
 \hline 
\end{tabular}
\caption{The DVCS cross section $\gamma^\ast p \rightarrow \gamma p$, $\sigma_{DVCS}$,
as a function of $W$ for  three $Q^2$ values.
The first errors are statistical, the second systematic.}
\label{sig2d}
\end{table}

\begin{table}[!htbp]
\centering
\begin{tabular}{|c|lcc|lcc|lcc|}
   \cline{2-10}
  \multicolumn{1}{c|}{~} &\multicolumn{9}{c|}{~} \\ [-10pt]
  \multicolumn{1}{l|}{} & \multicolumn{9}{c|}{$d\sigma_{DVCS}/dt
     \; \; \left[{\rm nb/GeV}^2\right]$} \\ [3.0pt]
  \cline{2-10}
   \multicolumn{1}{c|}{~} &\multicolumn{9}{|c|}{~} \\ [-11pt]
    \multicolumn{1}{c|}{~} &
   \multicolumn{9}{c|}{$W=82$~GeV} \\ [1.0pt]
\hline
  \multicolumn{1}{|c|}{~} &\multicolumn{3}{|c|}{~}
    &\multicolumn{3}{c|}{~}
    &\multicolumn{3}{c|}{~} \\ [-12pt]
  $|t|$ $\left[{\rm GeV}^2\right] $ &
  \multicolumn{3}{c|}{$Q^2 = $8$ \;$GeV$^2$} &  
  \multicolumn{3}{c|}{$Q^2 = $15.5$ \;$GeV$^2$} & 
  \multicolumn{3}{c|}{$Q^2 = $25$ \;$GeV$^2$} \\ [3.0pt]
 \hline 
       $0.10$&  $13.1$&  $\pm$	$1.10$&  $\pm$	 $1.85$&       $4.37$&    $\pm$  $0.47$&   $\pm$    $0.86$&    $1.41$&	$\pm$	 $0.40$&   $\pm$    $0.43$ \\
       $0.30$&   $4.69$&  $\pm$	$0.45$&  $\pm$	 $0.55$&       $1.02$&    $\pm$   $0.16$&   $\pm$    $0.18$&    $0.71$&	$\pm$	 $0.16$&   $\pm$    $0.08$ \\
       $0.50$&   $1.37$&  $\pm$	$0.21$&  $\pm$	 $0.23$&       $0.49$&    $\pm$   $0.08$&   $\pm$    $0.08$&    $0.28$&	$\pm$	 $0.07$&   $\pm$    $0.04$ \\
       $0.80$&   $0.19$&  $\pm$	$0.04$&  $\pm$	 $0.06$&       $0.12$&    $\pm$   $0.02$&   $\pm$    $0.02$&    $0.04$&	$\pm$	 $0.01$&   $\pm$    $0.02$ \\
 \hline 
 \hline 
  $b$ [GeV$^{-2}$]  & $5.84$ & $\pm$ $0.30$ & $\pm$ $0.35$ & $5.16$ & $\pm$ $0.26$ & $\pm$ $0.30$ & $5.09$ & $\pm$ $0.55$ & $\pm$ $0.60$  \\
 \hline 
%% \cline{2-10}
   \multicolumn{1}{c|}{~} &\multicolumn{9}{|c|}{~} \\ [-11pt]
    \multicolumn{1}{c|}{~} &
   \multicolumn{9}{c|}{$Q^2=10$~GeV$^2$} \\ [1.0pt]
\hline
  \multicolumn{1}{|c|}{~} &\multicolumn{3}{|c|}{~}
    &\multicolumn{3}{c|}{~}
    &\multicolumn{3}{c|}{~} \\ [-12pt]
  $|t|$ $\left[{\rm GeV}^2\right] $ &
   \multicolumn{3}{c|}{$W=40$ GeV} &
   \multicolumn{3}{c|}{$W=70$ GeV} &
   \multicolumn{3}{c|}{$W=100$ GeV} \\ [3.0pt]
 \hline
       $0.10$&    $4.99$&  $\pm$ $0.66$&  $\pm$  $0.54$&    $7.78$&  $\pm$  $0.69$& $\pm$ $0.87$&   $10.9$& $\pm$ $1.14$&$\pm$ $2.36$ \\
       $0.30$&    $1.45$&  $\pm$ $0.29$&  $\pm$  $0.18$&    $2.74$&  $\pm$  $0.31$& $\pm$ $0.30$&    $3.47$& $\pm$ $0.42$&$\pm$ $0.53$ \\
       $0.50$&    $0.49$&  $\pm$ $0.14$&  $\pm$  $0.08$&    $0.81$&  $\pm$  $0.14$& $\pm$ $0.11$&    $1.49$& $\pm$ $0.21$&$\pm$ $0.24$ \\
       $0.80$&   $0.12$&  $\pm$ $0.03$&  $\pm$  $0.03$&    $0.19$&  $\pm$  $0.03$& $\pm$ $0.03$&    $0.19$& $\pm$ $0.04$&$\pm$ $0.06$ \\
 \hline 
 \hline 
  $b$ [GeV$^{-2}$]  & $5.40$ & $\pm$ $0.40$ & $\pm$ $0.25$ & $5.34$ & $\pm$ $0.25$ & $\pm$ $0.27$ & $5.48$ & $\pm$ $0.31$ & $\pm$ $0.45$  \\
 \hline 
\end{tabular}
\caption{ The DVCS cross section $\gamma^* p\rightarrow \gamma p$, differential in $t$, $d\sigma_{DVCS}/dt$,
for three values of $Q^2$ at $W=82$~GeV, and 
for three values of $W$ at $Q^2=10$~GeV$^2$. Results for the corresponding $t$-slope parameters $b$ are given. 
The first errors are statistical, the second systematic.}
\label{sigtq}
\end{table}

%%%%%%%%%%%%%%%%%%%% geom scal

\newpage

\begin{table}[!htbp]
\centering
\begin{tabular}{|c|lcc|lcc|lcc|}
   \cline{2-10}
  \multicolumn{1}{c|}{~} &\multicolumn{9}{c|}{~} \\ [-10pt]
  \multicolumn{1}{l|}{} & \multicolumn{9}{c|}{$d\sigma_{DVCS}/dt
     \; \; \left[{\rm nb/GeV}^2\right]$} \\ [3.0pt]
  \cline{2-10}
   \multicolumn{1}{c|}{~} &\multicolumn{9}{|c|}{~} \\ [-11pt]
    \multicolumn{1}{c|}{~} &
   \multicolumn{9}{c|}{$Q^2=8$~GeV$^2$} \\ [1.0pt]
\hline
  \multicolumn{1}{|c|}{~} &\multicolumn{3}{|c|}{~}
    &\multicolumn{3}{c|}{~}
    &\multicolumn{3}{c|}{~} \\ [-12pt]
  $|t|$ $\left[{\rm GeV}^2\right] $ &
   \multicolumn{3}{c|}{$W=40$ GeV} &
   \multicolumn{3}{c|}{$W=70$ GeV} &
   \multicolumn{3}{c|}{$W=100$ GeV} \\ [3.0pt]
  \hline
       $0.10$&   $8.10$&  $\pm$ $1.22$&  $\pm$  $0.82$& $10.0$&  $\pm$  $1.30$&  $\pm$  $1.27$&  $16.0$&  $\pm$ $2.11$ & $\pm$	 $2.74$ \\
       $0.30$&   $2.30$&  $\pm$ $0.54$&  $\pm$  $0.28$&  $4.35$&  $\pm$  $0.63$&  $\pm$  $0.46$&   $5.45$&  $\pm$ $0.80$ & $\pm$	 $0.73$ \\
       $0.50$&   $0.45$&  $\pm$ $0.22$&  $\pm$  $0.10$&  $1.08$&  $\pm$  $0.27$&  $\pm$  $0.17$&   $1.96$&  $\pm$ $0.41$ & $\pm$	 $0.35$ \\
       $0.80$&   $0.16$&  $\pm$ $0.06$&  $\pm$  $0.03$&  $0.13$&  $\pm$  $0.06$&  $\pm$  $0.04$&   $0.21$&  $\pm$ $0.09$ & $\pm$	 $0.08$ \\
  \hline
%%  \cline{2-10}
%%  \multicolumn{1}{c}{~} &\multicolumn{9}{c}{~} \\ [-11pt]
%%  \cline{2-10}
  \multicolumn{1}{c|}{~} &\multicolumn{9}{|c|}{~} \\ [-11pt]
  \multicolumn{1}{c|}{~}  & \multicolumn{9}{c|}{$Q^2=20$~GeV$^2$} \\ [1.0pt]
\hline
  \multicolumn{1}{|c|}{~} &\multicolumn{3}{|c|}{~}
    &\multicolumn{3}{c|}{~}
    &\multicolumn{3}{c|}{~} \\ [-12pt]
  $|t|$ $\left[{\rm GeV}^2\right] $ &
   \multicolumn{3}{c|}{$W=40$ GeV} &
   \multicolumn{3}{c|}{$W=70$ GeV} &
   \multicolumn{3}{c|}{$W=100$ GeV} \\ [3.0pt]
 \hline 
       $0.10$&  $1.06$&   $\pm$ $0.28$&   $\pm$    $0.28$&	 $2.38$&   $\pm$    $0.29$&  $\pm$     $0.26$&       $2.98$&   $\pm$	$0.49$&	$\pm$	 $0.85$ \\
       $0.30$&  $0.33$&   $\pm$ $0.07$&   $\pm$    $0.07$&	 $0.67$&   $\pm$    $0.12$&  $\pm$     $0.07$&       $0.89$&   $\pm$	$0.17$&	$\pm$	 $0.17$ \\
       $0.50$&  $0.22$&   $\pm$ $0.06$&   $\pm$    $0.06$&	 $0.24$&   $\pm$    $0.05$&  $\pm$     $0.03$&       $0.44$&   $\pm$	$0.08$&	$\pm$	 $0.08$ \\
       $0.80$&  $0.04$&   $\pm$ $0.01$&   $\pm$    $0.01$&	 $0.07$&   $\pm$    $0.01$&  $\pm$     $0.02$&       $0.06$&   $\pm$	$0.02$&	$\pm$	 $0.02$ \\
 \hline 
\end{tabular}
\caption{ The DVCS cross section $\gamma^* p\rightarrow \gamma p$, differential in $t$, $d\sigma_{DVCS}/dt$,
for three values of $W$ extracted in two $Q^2$ intervals:  
$6.5 < Q^2 < 11$~GeV$^2$  and $11 < Q^2 < 80$~GeV$^2$, corrected to the central values of $Q^2=8$~GeV$^2$ and $20$~GeV$^2$, respectively.
The first errors are statistical, the second systematic.
}
\label{tabgs1}
\end{table}

%=========================================================================
\vfill
\newpage

%%%%%%%%%%%%%%%%%%%%%%%%%%%%%%%%%%%%%%%%%%%%%%%%%%

%%%%%%%%%%%%%%%%%%%%%%%%%%%%%%%%%%%%%%%%%%%%%%%%%%

%%%%%%%%%%%%%%%%%%%%%%%%%%%%%%%%%%%%%%%%%%%%%%%%%%

%%%%%%%%%%%%%%%%%%%%%%%%%%%%%%%%%%%%%%%%%%%%%%%%%%

%%%%%%%%%%%%%%%%%%%%%%%%%%%%%%%%%%%%%%%%%%%%%%%%%%

%=========================================================================

\end{document}